%% file: warpline_arxiv.tex
\documentclass{article}
\usepackage{microtype}
\usepackage{graphicx}
\usepackage{booktabs,array,multirow}
\usepackage{amsmath,amssymb}
\usepackage{xcolor,xspace}
\usepackage{hyperref}
\usepackage{xurl}

\usepackage[accepted]{mlsys2025}
\makeatletter
\renewcommand{\Notice@String}{Preprint.}
\makeatother
\mlsystitlerunning{Warpline}
\newcommand{\sys}{\textsc{Warpline}\xspace}
\newcolumntype{L}[1]{>{\raggedright\arraybackslash}p{#1}}
\begin{document}
\twocolumn[
\mlsystitle{Warpline: Binary-Level Tracing and Unified Evidence \\ for Agentic GPU Kernel Optimization}
\begin{mlsysauthorlist}
\mlsysauthor{Junyu Yin}{gmu}
\mlsysauthor{Vinod Grover}{nvidia}
\mlsysauthor{Keren Zhou}{gmu}
\end{mlsysauthorlist}
\mlsysaffiliation{gmu}{George Mason University}
\mlsysaffiliation{nvidia}{NVIDIA}
\mlsyscorrespondingauthor{Keren Zhou}{kzhou6@gmu.edu}
\mlsyskeywords{GPU kernel optimization, LLM coding agents, intra-kernel tracing, binary instrumentation, performance analysis, tensor cores}
\vskip 0.3in
\begin{abstract}
LLM agents are becoming GPU kernel engineers, but they still tune kernels without seeing where the time goes.
Hardware profilers report how busy each unit was, not where each warp's time went; intra-kernel tracers can show it, but only with hand-placed probes and a recompilation that alters the very kernel being measured.
We present \sys, which gives agents this view automatically while leaving the kernel essentially unchanged.
\sys analyzes the binary to decide which events to record and inserts probes into the binary itself without recompilation or any manual step.
It then aligns the trace with hardware profiling data on one unified timeline.
Because it needs only the binary, it can regenerate this evidence for every new version of the kernel at whatever level the agent edits it (source, IR, or PTX), closing the optimization loop without human intervention.
On NVIDIA B200, across 35 kernels in Triton, CuTe DSL, and CUDA C++, \sys adds a median overhead of 1.2\% and leaves a median 99.9\% of executed instructions unchanged, whereas existing tracers have median overheads of 1.6\% to 13.2\% and leave only 4.6\% to 32.7\% unchanged.
\sys also traces closed-source kernels, which these tracers cannot.
On 10 SOL-ExecBench problems, agents using \sys spend less time and fewer model requests than a bare coding agent and NVIDIA's Kernel Design Agents (KDA), yet lead both after every round in geometric mean and reach a 7.69$\times$ geometric-mean speedup, against 6.94$\times$ and 6.13$\times$.
Two case studies further demonstrate that \sys reveals hidden bottlenecks and provides actionable optimization insights.
\end{abstract}
]
\printAffiliationsAndNotice{}
\hypersetup{pdftitle={Warpline: Binary-Level Tracing and Unified Evidence for Agentic GPU Kernel Optimization},pdfauthor={Junyu Yin, Vinod Grover, Keren Zhou}}

\begin{figure*}[!htbp]
    \centering
    \includegraphics[width=\textwidth]{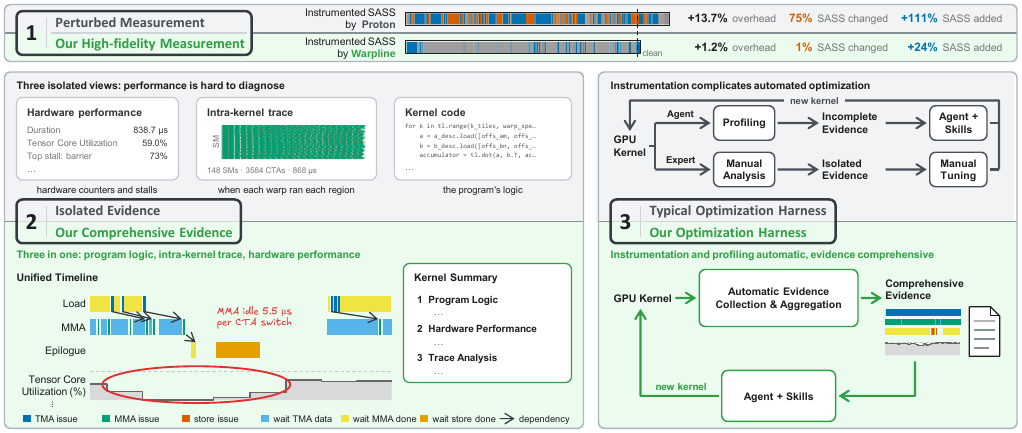}
    \caption{\sys{} versus standard practice on the Llama-3.1-8B MLP up-projection GEMM in Triton.
    (1)~Tracing with Proton changes 75\% of the original SASS instructions and adds 111\% more, with 13.7\% tracing overhead; \sys{} changes only 1\%, adds 24\%, and incurs 1.2\% overhead.
    (2)~Hardware profiles, traces, and code expose resource activity, execution timing, and program logic separately; \sys{}'s unified timeline shows tensor-core utilization dips coinciding with MMA idling at each CTA switch, and its kernel summary provides an overview.
    (3)~Standard practice can automate profiling but leaves instrumentation to experts; \sys{} automates instrumentation as well as profiling, so the agent receives comprehensive evidence in every round.}
    \label{fig:teaser}
\end{figure*}

\section{Introduction}
\label{sec:intro}

Developing high-performance tensor-core GPU kernels requires managing increasingly complex execution schedules~\cite{twill}.
Mechanisms such as warp specialization~\cite{cudadma,tawa}, software pipelining~\cite{alcop}, and asynchronous data movement and computation~\cite{cuda_async,flashattention3} enable greater overlap across warps and hardware units, but require coupled decisions about work partitioning, synchronization, buffering, and resource allocation. 
Coding agents~\cite{claudecode,codex} can help explore this design space~\cite{kernelbench,kda}, but need rich performance context and optimization guidance~\cite{kernelagent,kernelpro}.
Without them, agents may waste their budget on changes that do not address the bottleneck~\cite{kernelband}. 
Hardware profilers~\cite{nsight,nsightsystems} and intra-kernel tracers~\cite{proton,iket,neutrino} provide detailed but isolated views of kernel execution (Figure~\ref{fig:teaser}).
Using these observations together in an automated optimization loop presents three challenges.

\textbf{Low-overhead, high-fidelity intra-kernel tracing.}
A useful trace should represent the kernel's uninstrumented execution.
Yet every probe executes extra instructions to read the clock and store a record, and these instructions compete with the kernel for hardware resources.
Probes inserted into source code or IR~\cite{kperfir,protondocs} also participate in compilation and can change register allocation and instruction scheduling.
The resulting trace may therefore describe an altered implementation rather than the original kernel~\cite{xtrace}.
Effective intra-kernel tracing must minimize recording overhead while preserving the original kernel's execution behavior.

\textbf{Integrating isolated evidence.}
Intra-kernel tracers show which operations each warp executes and when, while hardware profilers report aggregate metrics and hardware activity over time.
These measurements come from separate runs, so aligning them on a unified timeline requires comparable execution across runs.
Tracing must therefore add little overhead and preserve the original kernel's behavior.
Connecting the two views also requires understanding the program logic to associate hardware activity with specific phases of each warp's execution.
These connections help agents distinguish different causes of low utilization and choose the right optimization target.

\textbf{Keeping the optimization loop automatic.}
As agents revise kernel implementations, performance evidence must remain both current and accessible.
Source-level probes are tied to the implementation's details, so structural rewrites may require an agent to inspect the new code and reposition probes.
The unified timeline contains millions of execution events with hardware activity sampled over time.
An image cannot expose this event-level detail, while reading the entire trace table is inefficient.
Diagnosing a slowdown requires inspecting specific operations, following their dependencies, and checking hardware activity over the relevant interval.
A fully automatic workflow therefore requires both instrumentation that adapts to code changes and queryable access to the resulting evidence.

We present \sys{}, a fully automated framework that provides comprehensive performance evidence for agentic optimization of tensor-core GPU kernels.
\sys{} traces kernels at the binary level.
It combines static and dynamic binary analysis to identify semantic events and inserts probes without source annotations or recompilation.
The probes keep the original instruction order, and their recording, register use, and scheduling are planned together to limit overhead.
Working on the binary also lets it support multiple programming languages and observe compatible closed-source kernels.
\sys{} then connects the traced events with hardware measurements and program logic in a unified timeline and a kernel summary.
Our skill package~\cite{agentskills} guides a coding agent to inspect this evidence and implement optimizations at the source, IR, or PTX~\cite{ptx} level.
An evaluator checks each candidate for correctness and performance, and every accepted kernel is traced again for the next round.

\begin{figure*}[!htbp]
    \centering
    \includegraphics[width=\textwidth]{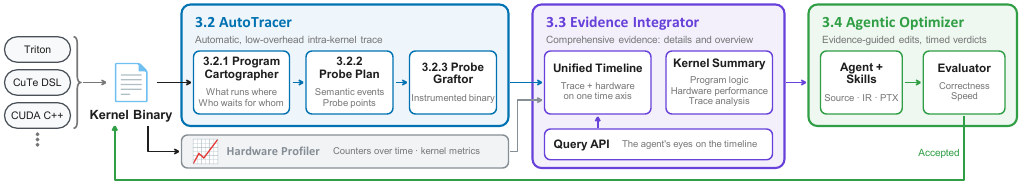}
    \caption{Overview of \sys{}. Each box is numbered by the subsection and grey marks the existing hardware profiler. An accepted candidate becomes the next round's kernel binary.}
    \label{fig:framework}
\end{figure*}

We evaluate our tracer on 35 kernels in Triton~\cite{tritonpaper}, CuTe DSL~\cite{cutedsl}, and CUDA C++~\cite{cudaguide}, plus three closed-source kernels, on NVIDIA B200~\cite{blackwell}.
Across the open-source kernels, it incurs 1.2\% median tracing overhead and preserves a median 99.9\% of the original instructions.
On 10 SOL-ExecBench~\cite{sol} problems, \sys{} maintains a lead throughout optimization at lower cost.
Its final kernels are 1.11$\times$ faster than a bare coding agent's and 1.25$\times$ faster than KDA's, in geometric mean.
It also improves three production kernels by 1.22--1.58$\times$. 
Case studies show how \sys{} uncovers hidden bottlenecks and translates the findings into code changes.

This paper makes three contributions:
\begin{itemize}
    \item An automatic tracer that uses binary analysis to identify semantic events and inserts low-overhead probes while preserving execution behavior, supporting multiple programming frontends and closed-source kernels.
    \item A comprehensive evidence interface that combines intra-kernel events, hardware measurements, and program logic in a unified timeline and kernel summary, with a query API for targeted analysis.
    \item An automated optimization harness that regenerates evidence as kernels evolve and uses a dedicated skill package to guide agents in interpreting evidence, editing source code, IR, or PTX, and evaluating candidates.
\end{itemize}

\section{Background}
\label{sec:background}

\subsection{Execution Model of Modern Tensor-Core Kernels}
\label{sec:bg-execution}

Modern tensor-core kernels coordinate asynchronous data movement and matrix computation across cooperating warps.
The \emph{Tensor Memory Accelerator} (TMA) transfers tiles between global and shared memory asynchronously, offloading address generation and data movement.
Matrix multiply--accumulate (MMA) instructions similarly submit computation without waiting for its completion.
On GPUs with \emph{tensor memory} (TMEM), accumulators can reside in dedicated on-chip storage.
Warp specialization (WS) assigns data loading, MMA issuance, and epilogue processing to different warp groups. 
Software pipelining overlaps their work across tiles: a multistage shared-memory ring lets TMA load future operands while tensor cores consume the current stage, and accumulator buffering can overlap an epilogue with the next tile's computation~\cite{cutlass_gemm}.
Because issuance does not imply completion, synchronization must establish both data readiness and safe buffer reuse.
Consumers wait for input transfers to complete; producers wait until consumers release the storage.
An \texttt{mbarrier} tracks arrivals and asynchronous transaction completion.
Repeated uses advance through phases, each completing once its expected arrivals and transactions have occurred~\cite{ptx}.

\subsection{GPU Compilation and Instrumentation}
\label{sec:bg-compilation}

Kernel compilers lower source programs through language-specific intermediate representations (IRs)~\cite{mlir} into PTX.
The PTX assembler (ptxas) selects instructions, allocates physical registers, and schedules instructions, producing executable SASS code packaged in a \texttt{cubin} binary~\cite{nvcc,binaryutils}.
Compilation can interleave operations from different source regions and introduce synchronization without explicit source statements.
Source- and IR-level probes express high-level regions but participate in subsequent compilation.
Binary instrumentation~\cite{nvbit} instead targets the final instructions without recompiling.
It must explicitly manage resources and instruction dependencies to preserve register values and control flow.
Because different programming frontends produce the same executable format, the binary provides a common instrumentation interface.
It also enables observation of closed-source kernels whose source and compiler IR are unavailable.

\section{Warpline}
\label{sec:method}

\subsection{Overview}
\label{sec:overview}

\sys{} is a fully automated framework that provides comprehensive performance evidence to help coding agents optimize tensor-core kernels more effectively (Figure~\ref{fig:framework}).
It works on the kernel binary, so it accepts kernels from any language that compiles to it.
From the captured binary, AutoTracer~(\S\ref{sec:autotracer}) automatically generates its intra-kernel trace by binary analysis, while the Hardware Profiler collects hardware performance measurements.
The Evidence Integrator~(\S\ref{sec:evidence}) then combines the trace, hardware measurements, and recovered program logic into comprehensive evidence as a unified timeline and a kernel summary.
We develop query APIs and a skill package for the Agentic Optimizer~(\S\ref{sec:optimizer}).
The skills guide coding agents in querying and interpreting evidence and using it to optimize kernels through edits at the source, IR, or PTX level.

\subsection{AutoTracer}
\label{sec:autotracer}

\subsubsection{Program Cartographer}
\label{sec:cartographer}

Conventional SIMT kernels assign the same computation to all warps, whereas WS kernels divide work among distinct warp groups.
Planning probes therefore requires identifying which groups execute which code and how they interact with each other.
We call these two aspects \emph{program logic}.
We recover this logic through abstract evaluation of SASS under the launch configuration.
\textbf{Warp groups and work loops} are identified by following each warp's control flow and grouping warps by reachable loops.
\textbf{Barrier topology} connects these groups. The evaluation tracks barrier initialization and operands to relate waits to contributions from arrivals, copies, and commits.

We do not model all kernel behavior statically: static analysis has inherent limits, and modeling shared memory and inter-warp communication is complicated.
Instead, we use an instrumented run to capture the shared-memory values used for warp dispatch and the register operands of synchronization instructions.
To preserve soundness, these observations are reused only when static checks prove that they depend solely on the kernel and launch configuration.
The run also records execution counts and lane participation, which guide the following Probe Graftor~(\S\ref{sec:probe-graftor}) to reduce tracing overhead.

\subsubsection{Probe Plan}
\label{sec:probe-plan}

The Probe Plan associates semantic events with their SASS instructions and the PCs at which probes record them.
We provide two recording levels here.
\emph{Lifetime} records each warp's start and end to reveal work distribution, execution overlap, and completion tails.
\emph{Pipeline} adds work loops and events for data movement, tensor-core operations, and synchronization.
These events show how cooperating warps progress through the pipeline.
Events are recorded per warp as \emph{spans} or \emph{instants}.
A span brackets a region with start and end timestamps, such as a warp waiting at a barrier for its input data.
An instant records the time just after an instruction.
For issue and commit instructions, this timestamp marks submission rather than completion of asynchronous work.

\begin{figure*}[!htbp]
    \centering
    \includegraphics[width=\textwidth]{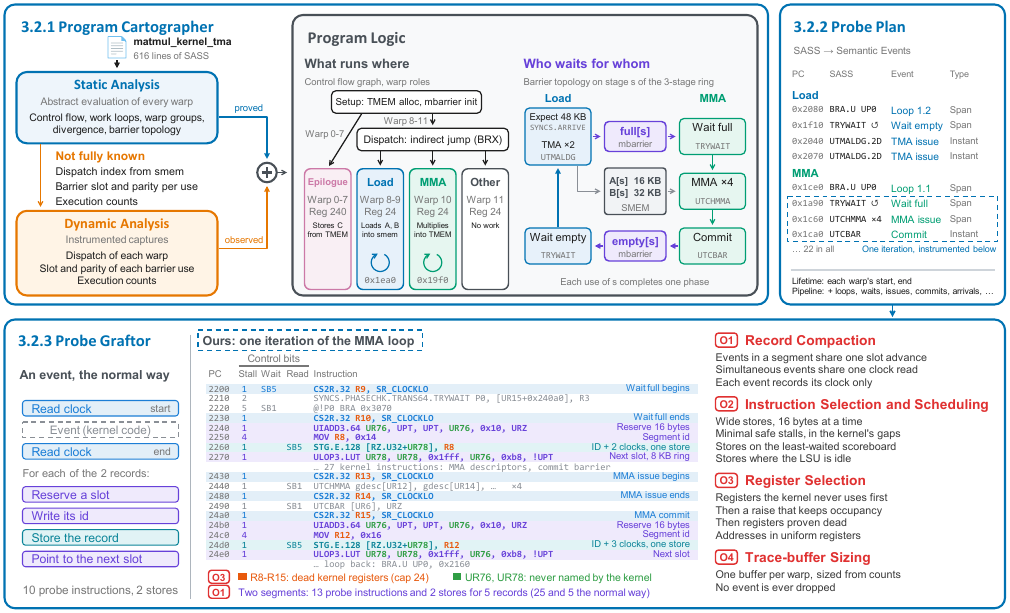}
    \caption{AutoTracer on the Llama-3.1-8B~\cite{llama3} MLP up-projection GEMM kernel in Triton.
    \textbf{Top}: binary analysis yields the program logic, and the Probe Plan maps SASS instructions to semantic events.
    \textbf{Bottom}: one iteration of the MMA loop as instrumented by our Probe Graftor, against conventional per-event recording.
    Its five records take only 13 probe instructions, including 2 stores, instead of 25 including 5.}
    \label{fig:autotracer}
\end{figure*}

\subsubsection{Probe Graftor}
\label{sec:probe-graftor}

Given the Probe Plan, the Probe Graftor inserts probes directly into the kernel binary without recompilation to preserve the original instructions and their order.
The full plan and binary analysis let us coordinate recording, register use, and scheduling across probes.
The following four optimizations exploit this information to reduce tracing overhead.

\paragraph{O1: Record compaction.}
Beyond reading the clock, a probe requires additional SASS instructions to reserve buffer space, write an identifier, and advance the pointer.
Our analysis groups probe points that a converged warp always executes into a \emph{segment}, performing these instructions only once per segment.
Each probe then records only its timestamp, with its identity determined statically by its position within the segment.
When two events share a boundary, their clock reads are also merged.

\paragraph{O2: Instruction selection and scheduling.}
Probes compete with the kernel for issue slots, execution pipelines, and scoreboards~\cite{moderncores}.
We pack each segment's records into wide stores of up to 16 bytes, reducing the number of store instructions issued.
Because stores need not coincide with event boundaries, each may move to a later point within its segment where the uninstrumented kernel's stall samples indicate low memory pressure or existing warp stalls.
Every inserted instruction takes the smallest stall count its dependencies allow, and each store is tracked on a scoreboard the kernel never uses, or on the one it waits on least often.

\paragraph{O3: Register selection.}
The Probe Graftor first uses registers unused by the kernel, then raises the declared count without reducing occupancy, and finally reuses registers proved dead by liveness analysis~\cite{xtrace}.
If none are available, it borrows live registers, saving their values to local memory and restoring them after recording.
Per-warp recording allows buffer addresses and offsets to use uniform registers, avoiding additional general-purpose register pressure.

\paragraph{O4: Trace-buffer sizing.}
Recording demand varies across warps, so uniformly sized buffers can waste global memory or wrap around and overwrite earlier events.
Here we estimate an upper bound on per-warp demand from the dynamic run to retain all events without excessive allocation.

\subsubsection{A Running GEMM Example}
\label{sec:gemm-example}

Figure~\ref{fig:autotracer} illustrates AutoTracer on the Triton GEMM for Llama-3.1-8B's MLP up projection.
Static analysis splits the 12 warps into 2 groups.
Warps 0--7 branch on their warp index to the epilogue, under a 240-register limit.
Warps 8--11, under a 24-register limit, reach an indirect jump (\texttt{BRX}) whose index comes from shared memory.
The dynamic run further captures the target each warp takes: warps 8--9 the load loop, warp 10 the MMA loop, and warp 11 code with no work loop.
Static analysis also recovers the three-stage shared-memory ring between Load and MMA.
TMA completion signals \texttt{full[s]} for MMA, while MMA's committed arrival signals \texttt{empty[s]} for Load.
The dynamic run here captures each wait's stage and parity.
The Probe Plan selects 22 events automatically.
For example, the \texttt{TRYWAIT} at \texttt{0x1a90} and its retry branch form a span where MMA waits on \texttt{full[s]} for input tiles, while the \texttt{UTMALDG} at \texttt{0x2040} marks a TMA-issue instant.
The bottom shows one iteration of the MMA loop instrumented by the Probe Graftor.
The iteration is cut into two segments: the wait, with its begin and end, and the MMA issue with its commit.
Each segment uses \texttt{UIADD3}, \texttt{MOV}, and \texttt{ULOP3} once for reservation, identification, and address advancement, respectively.
Each event boundary records only its time, with one \texttt{CS2R.32}.
A single 16-byte \texttt{STG.E.128} store is used once enough data have accumulated.
Probes use dead registers \texttt{R8}--\texttt{R15} under the 24-register limit and unused uniform registers \texttt{UR76} and \texttt{UR78}.

\subsection{Evidence Integrator}
\label{sec:evidence}

\paragraph{Hardware performance.}
We use Nsight Compute (NCU) to collect hardware performance metrics, including 19 performance-monitor (PM) sampling time series.
They can be divided into four groups.
\emph{Compute \& Scheduler} (4 series) covers tensor-core activity, issue activity, and the active and eligible warps per SM.
\emph{Memory System} (6 series) covers tensor-memory activity, TMA request activity, shared-memory wavefronts and bank conflicts, and L2 and DRAM throughput.
\emph{Warp Stalls} (6 series) counts the warps stalled on long and short scoreboards, on the local/global and MIO queue throttles, at barriers, and at memory barriers.
\emph{Runtime Resources} (3 series) tracks the registers, shared memory, and tensor memory allocated per SM.

\paragraph{Unified timeline.}
Each warp group is assigned a role based on the operations within its recorded work loops.
Dependencies are recovered per barrier slot: each wait is paired with the arrivals, copies, and commits that complete the phase it observed.
We keep a pairing only where the arrivals counted in the trace prove which phase that was.

Hardware measurements come from a separate run and so require alignment with the trace.
We fit an affine mapping to align warp lifetime in the trace with the profiler's active-warp series, then apply it to remaining hardware series.
Because the traced run executes nearly the same instructions at nearly the same speed, the aligned series describe the traced execution.

\paragraph{Kernel summary.}
The kernel summary combines program logic, hardware measurements, and trace analysis to provide an overview of the execution.
The analysis shows where each role spends time and whether work is balanced across SMs.
Using the recovered barrier topology, it also provides blame-shifting analysis~\cite{hpctoolkitgpu}.
For example, a hotspot analysis may flag a long MMA wait, while blame shifting reveals a late TMA submission by Load.

\paragraph{Query API.}
Agents cannot easily extract detailed execution information from the timeline.
Our query API provides access to the underlying events and hardware measurements, allowing agents to ask questions like how hardware activity changes during a particular phase.
An agent can thus start from a finding in the summary and check it against the events behind it.

\subsection{Agentic Optimizer}
\label{sec:optimizer}

\paragraph{Skill-guided optimization.}
Our skill package guides the agent from interpreting evidence to implementing and evaluating optimizations.
The skills explain how to read the kernel summary, investigate execution details through the query API, and use hardware knowledge to form testable hypotheses.
They also provide concrete instructions for turning optimization ideas into edits to source code, compiler IR, or PTX and using our evaluator to assess the results.

\paragraph{Validation and timing.}
The evaluator checks kernel correctness on multiple inputs and measures latency using Hardware Event System through CUPTI~\cite{cupti}.
Parent and candidate measurements use ABBA ordering, with each round starting with warm-up and collecting medians from repeated kernel invocations, each preceded by an L2 flush.
We compare geometric means of batch medians.
A correct candidate is accepted only if its relative improvement exceeds the parent's half-band, which is half the parent's latency range divided by its median.

\section{Evaluation}
\label{sec:evaluation}

\subsection{Experimental Setup}
\label{sec:setup}

\paragraph{Environment and kernels.}
We evaluate \sys{} on kernels written in Triton~\cite{triton}, CuTe DSL~\cite{cutlass}, and CUDA C++, using NVIDIA B200 GPUs (148 SMs, \texttt{sm\_100a}).
Table~\ref{tab:setup} lists the software, the profilers compared, the agent models, and each study's kernels by source and language.

\begin{table}[t]
\caption{Experimental environment and kernels. The 10 SOL-ExecBench problems~\cite{sol} cover convolution, recurrence and scan, attention, quantized and dense GEMM, two each.
The production and tutorial kernels come from the Triton repository. The cuBLAS and cuDNN kernels are closed source.}
\label{tab:setup}
\centering\small
\setlength{\tabcolsep}{4pt}
\begin{tabular}{@{}L{30mm}>{\centering\arraybackslash}p{10mm}>{\centering\arraybackslash}p{14.5mm}>{\centering\arraybackslash}p{17mm}@{}}
\toprule
Component & \multicolumn{3}{l@{}}{Configuration} \\
\midrule
GPU / target & \multicolumn{3}{L{\dimexpr41.5mm+4\tabcolsep\relax}@{}}{NVIDIA B200 / \texttt{sm\_100a}} \\
CUDA driver / toolkit & \multicolumn{3}{L{\dimexpr41.5mm+4\tabcolsep\relax}@{}}{595.58.03 / 13.2} \\
Nsight Compute & \multicolumn{3}{L{\dimexpr41.5mm+4\tabcolsep\relax}@{}}{2026.1.1} \\
CUPTI & \multicolumn{3}{L{\dimexpr41.5mm+4\tabcolsep\relax}@{}}{13.0.85} \\
Triton / CuTe DSL & \multicolumn{3}{L{\dimexpr41.5mm+4\tabcolsep\relax}@{}}{3.8.0 / 4.6.1} \\
Proton / IKET & \multicolumn{3}{L{\dimexpr41.5mm+4\tabcolsep\relax}@{}}{with Triton 3.8.0 / CuTe DSL 4.6.1} \\
FlashInfer / Neutrino & \multicolumn{3}{L{\dimexpr41.5mm+4\tabcolsep\relax}@{}}{\texttt{9c68e44} / \texttt{4a82cd2}} \\
Agent model & \multicolumn{3}{L{\dimexpr41.5mm+4\tabcolsep\relax}@{}}{Claude Sonnet 5.5, effort \texttt{xhigh}} \\
KDA reviewer & \multicolumn{3}{L{\dimexpr41.5mm+4\tabcolsep\relax}@{}}{GPT-6.1-sol, effort \texttt{xhigh}} \\
\midrule
\midrule
Kernel source & Triton & CuTe DSL & CUDA C++ \\
\midrule
\multicolumn{4}{@{}l}{\emph{Profiling}} \\
SOL-ExecBench & 10 & 10 & 10 \\
Production kernels & 3 & -- & -- \\
Tutorial kernels & 2 & -- & -- \\
cuBLAS & \multicolumn{3}{c@{}}{nvJet GEMM} \\
cuDNN & \multicolumn{3}{c@{}}{Flash-attention forward (SDPA)} \\
cuDNN & \multicolumn{3}{c@{}}{Implicit-GEMM convolution} \\
\midrule
\multicolumn{4}{@{}l}{\emph{Optimization}} \\
SOL-ExecBench & 10 & 3 & 3 \\
Production kernels & 3 & -- & -- \\
\bottomrule
\end{tabular}
\end{table}

\subsection{Tracing Overhead and Fidelity}
\label{sec:profiler-eval}

\begin{figure*}[t]
\centering
\includegraphics[width=\textwidth]{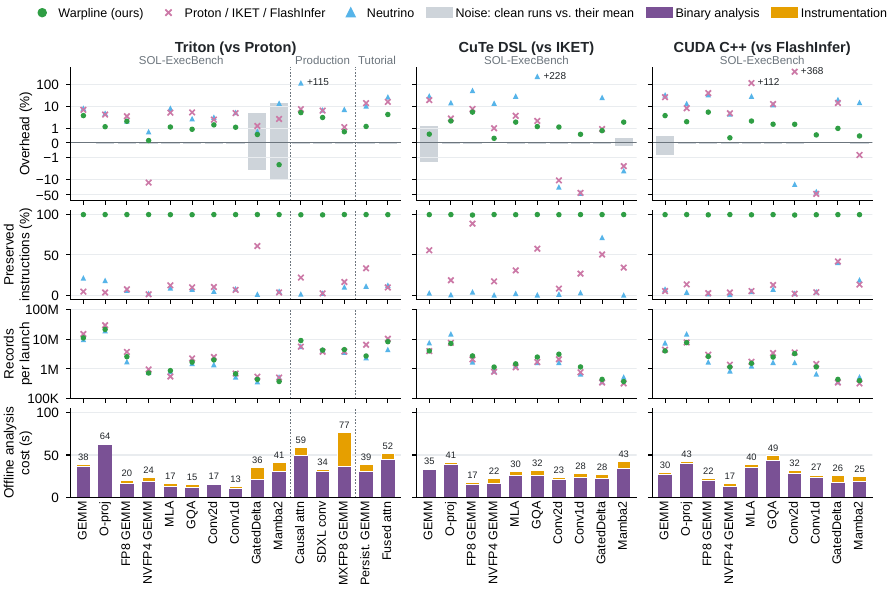}
\caption{Tracing the 35 open-source kernels at the Pipeline level with AutoTracer, each language's own tracer, and Neutrino.
\textbf{Rows, top to bottom}: runtime overhead against the uninstrumented binary; instruction preservation ratio; the number of trace records per launch; and AutoTracer's analysis and instrumentation time.
Grey bands: noise across repeated runs of the uninstrumented binary.}
\label{fig:tracing}
\end{figure*}

AutoTracer operates on kernel binaries and is therefore independent of source language.
We evaluate its overhead and fidelity against Proton~\cite{proton} for Triton, IKET~\cite{iket} for CuTe DSL, FlashInfer's profiler~\cite{flashinfer} for CUDA C++, and Neutrino~\cite{neutrino}, a PTX-level tracer supporting all three.
The evaluation covers the 38 kernels in Table~\ref{tab:setup}: 35 written in these languages and three closed-source library kernels, with comparisons restricted to each tracer's supported kernels.

For the four baselines, we manually place probes at the Pipeline-level events of \S\ref{sec:probe-plan} wherever supported.
Figure~\ref{fig:tracing} (third row) reports record counts per launch.
AutoTracer records 0.4--2.0 times as many events as the corresponding baseline, with per-tracer median ratios of 0.8--1.3, and records more events in 41 of 70 comparisons.
These results support two observations: (1) automatic planning provides coverage comparable to manual instrumentation and (2) its overhead and fidelity advantages cannot be explained by systematically recording fewer events.

\begin{figure*}[!htbp]
\centering
\includegraphics[width=\textwidth]{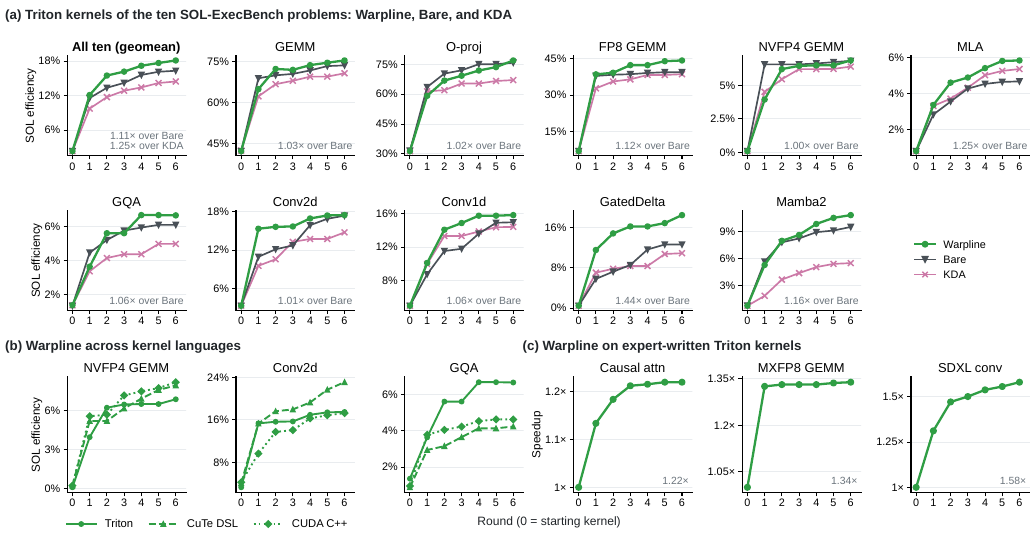}
\caption{Kernel optimization over six rounds. (a) SOL efficiency on ten SOL-ExecBench problems implemented in Triton, with the geometric mean across problems shown first.
(b) SOL efficiency of \sys{} across Triton, CuTe DSL, and CUDA C++ implementations of three problems.
(c) Speedup over the original implementation for three expert-written production kernels.}
\label{fig:optimization}
\end{figure*}

\paragraph{Overhead.}
We time each kernel's uninstrumented and instrumented binaries with the evaluator's high-precision timer (\S\ref{sec:optimizer}).
For uninstrumented latency $T_0$ and instrumented latency $T_1$, overhead is $(T_1-T_0) / T_0 \times 100\%$.
IKET holds the CUPTI interface required by this timer, so we obtain its instrumented latency from the kernel duration in its trace.
We report medians and ranges, because the other tracers' overheads have outliers in both directions.
AutoTracer's overhead is low on every kernel (Figure~\ref{fig:tracing}, top): it ranges from $-$2.1\% to 5.4\%, with a median of 1.2\%. 
The other tracers cost more and vary widely.
Proton's overhead ranges from $-$13.8\% to 15.7\% (median 4.8\%), IKET's from $-$40.5\% to 19.0\% (median 1.6\%), FlashInfer's from $-$45.0\% to 368\% (median 13.2\%), and Neutrino's from $-$41.6\% to 228\% (median 10.3\%).
AutoTracer’s overhead is $-2.1\%$ on Triton's Mamba2 kernel, whose execution time fluctuates substantially: repeated uninstrumented runs vary from $-10\%$ to $+14\%$ around their mean.
The other tracers also show several negative overheads, but these arise from changes to the program, as shown below.

\paragraph{Fidelity.}
An original instruction counts as unchanged if the instrumented binary keeps its encoding and its order among the other original instructions.
Branches whose targets shift only because code was inserted also count as unchanged.
Instruction preservation is the fraction of unchanged instructions weighted by how often each one executes in the uninstrumented kernel (Figure~\ref{fig:tracing}, second row).
Proton's preservation ranges from 1.3\% to 61.1\% (median 9.9\%), IKET's from 8.4\% to 88.7\% (median 32.7\%), FlashInfer's from 2.1\% to 42.0\% (median 5.4\%), and Neutrino's from 0.6\% to 71.7\% (median 4.6\%).
One Proton build (the persistent GEMM) spills registers where the uninstrumented kernel does not.
These tracers therefore trace a different program from the one they are meant to measure.
Some of their builds even run more than 10\% faster than the uninstrumented kernel with a preservation below 27\%.
AutoTracer inserts its probes into the compiled binary and leaves almost all original instructions untouched.
Its preservation ranges from 99.4\% to 100\% with a median of 99.9\%.

\paragraph{Closed-source kernels.}
AutoTracer also traces closed-source kernels.
Overhead is 0.21\% for cuBLAS nvJet GEMM, 3.07\% for cuDNN SDPA, and 0.32\% for cuDNN convolution, with instruction preservation 99.85--99.99\%.
The compared tracers require source code or PTX and therefore cannot trace them.

\paragraph{Offline analysis cost.}
To get an instrumented kernel, AutoTracer analyzes the binary and inserts probes automatically.
This takes 13 to 77\,s per kernel (median 30\,s; Figure~\ref{fig:tracing}, bottom): the analysis takes 11 to 62\,s (median 26\,s) and the instrumentation 2 to 40\,s (median 4\,s).
The median is under 1\% of a 60-minute optimization round, and no one has to read the code or place probes.

\begin{table}[!htbp]
\caption{Optimization outcomes and agent activity on the ten problems.
Speedup and SOL efficiency are geometric means across problems.
Differences within 3\% of Bare are counted as ties.
KDA calls are shown as coding-agent calls plus reviewer calls.}
\label{tab:arms}
\centering\small
\setlength{\tabcolsep}{3pt}
\begin{tabular}{@{}lccc@{}}
\toprule
 & \sys{} & Bare & KDA \\
\midrule
Speedup & \textbf{7.69$\times$} & 6.94$\times$ & 6.13$\times$ \\
SOL efficiency (\%) & \textbf{18.1} & 16.3 & 14.5 \\
Wins/ties/losses vs.\ Bare & 6/4/0 & -- & 1/1/8 \\
\midrule
Agent minutes / round & \textbf{30.7} & 33.5 & 59.9 \\
Model requests / round & \textbf{44} & 54 & 69 + 39 \\
Summary reads / round & 7.0 & -- & -- \\
Trace queries / round & 2.2 & -- & -- \\
NCU runs / round & 0.22 & 0.75 & 2.57 \\
\bottomrule
\end{tabular}
\end{table}

\subsection{Optimization Quality and Efficiency}
\label{sec:optimization-eval}

We compare \sys{} with a bare agent and KDA~\cite{kda} on ten SOL-ExecBench problems implemented in Triton to evaluate whether it improves optimization quality and efficiency under a fixed budget.
All three use the same prompt template and six rounds of up to 60 minutes each, with each round's candidate evaluated by the same evaluator.
The coding agent is Claude Sonnet 5.5 (\texttt{xhigh}); KDA additionally uses GPT-6.1-sol (\texttt{xhigh}) for review.
Each SOL-ExecBench problem contains multiple workloads, and we use their geometric-mean latency as the problem's latency.
Because starting kernels differ across problems, we report \emph{SOL efficiency}, defined as $\operatorname{GM}_{w}(T_{\mathrm{SOL},w}/T_w)$, where $T_w$ is the measured latency for workload $w$ and $T_{\mathrm{SOL},w}$ is the benchmark's speed-of-light latency bound on B200.

\paragraph{Optimization quality and progress.}
\sys{} achieves a 7.69$\times$ speedup over the starting kernels and 18.1\% SOL efficiency, compared with 6.94$\times$ and 16.3\% for Bare and 6.13$\times$ and 14.5\% for KDA (Table~\ref{tab:arms}).
The advantage also appears earlier (Figure~\ref{fig:optimization}a): after two rounds, \sys{} surpasses KDA's six-round result; after four rounds, it surpasses Bare's six-round result.
Figure~\ref{fig:optimization}a also details each problem's progress: \sys{} finishes with six wins and four ties against Bare and ten wins against KDA, maintaining its lead over both from round 1 onward on most problems.

\paragraph{Agent cost and evidence use.}
\sys{} also achieves lower agent cost (Table~\ref{tab:arms}).
It averages 30.7 minutes and 44 model requests per round, compared with 33.5 minutes and 54 requests for Bare, and 59.9 minutes and 108 requests for KDA.
The agent uses the supplied evidence, reading the kernel summary 7.0 times and querying the trace 2.2 times per round.

\paragraph{Other languages and production kernels.}
\sys{} improves SOL efficiency across the CuTe DSL and CUDA C++ implementations of all three problems, extending the gains observed in Triton (Figure~\ref{fig:optimization}b).
For the expert-written production kernels in Triton, Figure~\ref{fig:optimization}c reports speedups over the original implementations: 1.22$\times$ on causal attention, 1.34$\times$ on MXFP8 GEMM, and 1.58$\times$ on SDXL convolution.
\sys{} thus works across languages and still finds speedups in kernels that experts have tuned.

\subsection{Case Study 1: Bunched Epilogue Stores}
\label{sec:case1}

\begin{figure}[t]
\centering
\includegraphics[width=\columnwidth]{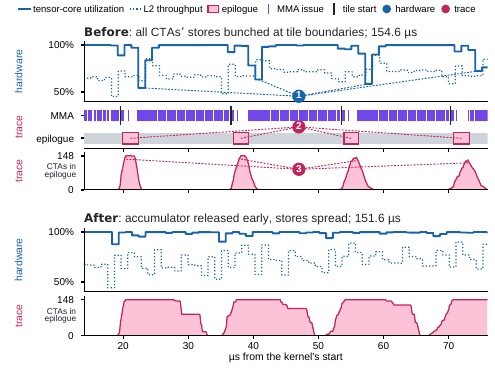}
\caption{Unified timelines of SDXL convolution before and after round 3.
Hardware tracks show tensor-core utilization and L2 throughput from the uninstrumented kernel's PM samples, aligned to the trace.
Trace tracks show one CTA's MMA and epilogue warps and the number of CTAs in epilogue.
(1)~Utilization drops and L2 throughput rises every 17\,\textmu s.
(2)~Within one CTA, the MMA warp issues few MMAs during each epilogue.
(3)~The drops coincide with nearly all 148 CTAs executing their epilogues.
After the change, each epilogue spreads stores over about 11\,\textmu s.}
\label{fig:case-epilogue}
\end{figure}

This case comes from the third optimization round of SDXL convolution, an expert-written production kernel in Triton (Figure~\ref{fig:optimization}c).
Looking only at the hardware profile, we observe 86\% average tensor-core utilization, with dips to 55--75\% lasting about 2\,\textmu s every 17\,\textmu s.
These dips coincide with increased L2 throughput (Figure~\ref{fig:case-epilogue}, marker~1).
This pattern suggests operand starvation, motivating a deeper operand ring or a tile order with better L2 reuse.
Yet the agent had already tested both in the previous round: a ninth ring stage left latency unchanged, and a different tile order changed it by less than 0.4\%.
The hardware profile thus points toward changes that fail to address the bottleneck.
The trace alone shows that the epilogue slows the MMA warp, but not why.
In each CTA, a tile's epilogue takes about 2.4\,\textmu s and overlaps the MMA warp's next tile.
During it, the MMA warp issues about one MMA per microsecond instead of six (marker~2), although it waits neither for the epilogue (0.2\% of its life) nor for its operands.
Read alone, the trace shows no wait that explains the slowdown.

With \sys{}, the agent diagnoses the drops by reading the hardware samples and the trace together.
On the trace's time axis, every drop falls where most CTAs are in their epilogue (marker~3).
The CTAs run nearly in lockstep, so every MMA warp slows at once, and the whole GPU writes its output tiles in one burst, which is when L2 throughput rises.
The agent therefore attributes the drops to the bunched epilogue stores, not to the operand loads, which the MMA warp does not wait for.
The agent changed the epilogue to read the accumulator from tensor memory into registers, release it at once, and spread its four sub-tile stores over about 11\,\textmu s with short sleeps (After).
The MMA warp then issues as fast during an epilogue as outside it, utilization stays at or above 88\%, and the kernel runs in 151.6\,\textmu s instead of 154.6\,\textmu s.

\subsection{Case Study 2: Compiler-Inserted Cluster Barriers}
\label{sec:case2}

\begin{figure}[!htbp]
\centering
\includegraphics[width=\columnwidth]{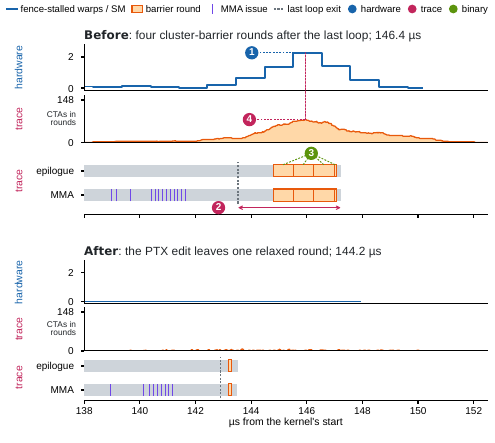}
\caption{The SDXL convolution's last microseconds before and after round 6: fence-stalled warps (hardware), CTAs in barrier rounds, and one CTA's warps (trace).
(1)~Fence stalls at the kernel's end.
(2)~3.6\,\textmu s from the last loop (dashed) to exit.
(3)~Four compiler-inserted cluster-barrier rounds, each after a GPU-wide fence.
(4)~The fence stalls track the CTAs in these rounds.
After the PTX edit, one relaxed round remains.}
\label{fig:case-teardown}
\end{figure}

This case comes from round 6 of the same SDXL convolution, which runs in 146.4\,\textmu s with 91\% average tensor-core utilization.
Existing tools, even used together, cannot locate the cost left at the kernel's end.
A hardware profiler sees fence-stalled warps rising from 0.08 to about 2 per SM as the kernel completes (Figure~\ref{fig:case-teardown}, marker~1).
But none of the stalled instructions ranks among the hottest, and the line table maps them to the kernel's definition line, so the profile names no code to blame.
A source-level tracer cannot see these microseconds at all.
Its probes reach no further than the epilogue's own code, which ends with the loop and the barrier invalidations after it.
The CTA exits 3.6\,\textmu s after the loop ends (marker~2), but nothing after those invalidations has a statement to probe, so the tracer's timeline stops there.

\sys{} sees these microseconds because AutoTracer probes the final binary, not the source.
About 1.2\,\textmu s after the last loop, all eight warps of both CTAs in the cluster run four rounds of cluster arrive and wait, about 0.7\,\textmu s each (marker~3).
Triton's 2-CTA lowering inserts these rounds into the PTX when the warp-specialized partitions end, and ptxas compiles each non-relaxed \texttt{barrier.cluster.arrive} into a GPU-wide memory fence (\texttt{MEMBAR.ALL.GPU}) followed by the cluster barrier.
On the unified timeline, the profiler's fence stalls track the number of CTAs in these rounds, which peaks at 80 of the 148 (marker~4).
Apart from the barrier invalidations at its start, the code in this interval exists in neither the source nor the Triton IR, so only a PTX edit can change it.
Guided by our trace and skills, the agent edits the PTX: it deletes three of the four rounds, relaxes the remaining one, and removes the barrier invalidations before exit.
The time from the last loop to exit falls from 3.6\,\textmu s to 0.46\,\textmu s, and the kernel runs in 144.2\,\textmu s instead of 146.4\,\textmu s.

\section{Related Work}
\label{sec:related}

\subsection{GPU Performance Analysis}

Performance models~\cite{roofline,hierarchicalroofline} and hardware profilers~\cite{nsight} characterize resource efficiency by relating computation, data movement, and measured utilization to hardware limits.
Application-level and context-sensitive profilers~\cite{nsightsystems,hpctoolkitgpu,contextprofile} associate execution costs with program regions and calling contexts, exposing expensive code paths.
Diagnostic analyses~\cite{gpa,leo,valueexpert} combine measurements with instruction dependencies or value patterns to identify stall sources and specific inefficiencies.
\sys complements these tools by correlating intra-kernel events, hardware-counter time series, and program logic through a unified timeline and a kernel summary, relating hardware utilization changes to the progress and dependencies of cooperating warps.

\subsection{Intra-Kernel Tracing and Instrumentation}

Source- and IR-level methods~\cite{nanotrace,cudaadvisor,cudaflux,kperfir,neutrino,proton} insert probes during compilation, potentially altering compiler optimizations and kernel execution behavior.
IKET~\cite{iket} reserves resources and placeholders during compilation and replaces the placeholders with probes afterward.
At the SASS level, SASSI and NVBit~\cite{sassi,nvbit} inject calls to probe routines, CUTracer~\cite{cutracer} builds instruction analyses on NVBit, and Xtrace~\cite{xtrace} directly splices probes into compiled kernels to reduce tracing overhead.
\sys automatically identifies semantic events and inserts low-overhead probes into kernel binaries without source annotations or recompilation.

\subsection{Automated GPU Kernel Optimization}

Compiler-based approaches~\cite{tvm,ansor,mirage,prism} search tensor programs and schedules using structured representations, cost models, and symbolic reasoning.
Specialized compilers~\cite{alcop,tawa,twill} automate software pipelining and warp specialization to overlap computation and data movement.
Learning-based approaches~\cite{kevin,cudal1,cudaagent} train code-generation models with correctness and performance feedback to improve iterative kernel optimization.
Agentic workflows~\cite{kda,kernelskill,kernelband,kernelagent,keet,tritonforge,kernelpro} combine reusable knowledge, search strategies, and profiling-based explanations to guide kernel generation.
\sys automatically regenerates comprehensive performance evidence as kernels evolve and exposes it through queries and skills to guide coding agents in a fully automated optimization loop.

\section{Conclusion}
\label{sec:conclusion}

We presented \sys, which automatically gives coding agents comprehensive evidence of how tensor-core kernels execute, relating the progress of each warp to the activity of the hardware over time.
On the kernels we evaluated, its tracing adds a median overhead of 1.2\% and leaves at least 99.4\% of executed instructions unchanged.
On ten SOL-ExecBench problems, agents using \sys produce final kernels 1.11$\times$ and 1.25$\times$ faster than a bare coding agent and KDA, in geometric mean.
Extending AutoTracer beyond \texttt{sm\_100a} is future work.

\label{END-OF-BODY}
\bibliographystyle{mlsys2025}
\bibliography{refs_verified}

\section*{AI Usage Disclosure}
We used AI tools, including Claude Code and ChatGPT, during the development of \sys{} and preparation of this manuscript.
The authors determined the research questions, system design, and experimental protocol, and reviewed and validated all AI-assisted work.
Kernels produced by the LLM agents evaluated in our optimization study were checked by our deterministic evaluator. 
The authors take full responsibility for the content of this paper.

\onecolumn
\appendix
\input{appendix}

\end{document}

%% file: appendix.tex
\makeatletter\setlength{\@fptop}{0pt}\makeatother
\raggedbottom
\makeatletter
\newcommand{\appshot}[4][\textwidth]{\par\medskip\noindent\begin{minipage}{\textwidth}\centering\includegraphics[width=#1]{figs/appendix/#2}\def\@captype{figure}\caption{#3}\label{#4}\end{minipage}\par\medskip}
\makeatother

\section{Reading the Unified Timeline}
\label{app:timeline}

Each figure in this section is a screenshot of a \sys{} trace in the Perfetto UI~\cite{perfetto}, cropped to the tracks we discuss, with a legend of the instant events added.

\paragraph{Layout.}
Hardware tracks come first.
They are the PM series of \S\ref{sec:evidence} (Appendix~\ref{app:pm}).
Figure~\ref{fig:app-panorama} shows all 19.
The other figures pin some of them above the warps.
The trace follows as a tree of kernel, SM, CTA slot, role, and warp.
A slot is one place for a CTA on an SM, so the CTAs that run on an SM one after another share a slot.
A role track carries its name, its warps, and its register limit, for example \texttt{load:warps 8-9, MAXREG 24}.
A warp's spans nest: its lifetime (\texttt{warp}), its work loops, and the events inside them.
An instant has no duration and appears as a chevron.

\paragraph{Names.}
Each name says whose event it is, so a producer's name matches the waits it ends.
A \texttt{TMA load} completes a \texttt{wait TMA}, an \texttt{MMA commit} completes a \texttt{wait MMA}, and a \texttt{softmax2 arrive} completes a \texttt{wait softmax2}.

\paragraph{Arrows and arguments.}
Selecting a slice shows its arguments (Table~\ref{tab:app-args}) in the details panel and draws its dependency arrows.
An arrow starts at a contribution (a copy, a commit, or an arrival) and ends where the wait it completed returned.

\paragraph{The last nanosecond of a wait.}
An arrow into a wait should end where the wait completes, but Perfetto attaches it to the start of the slice.
We therefore split the last nanosecond off each such wait as a separate slice with the same name and arguments, and end the arrow on it.
Figure~\ref{fig:app-lastns} shows one such wait.

\appshot{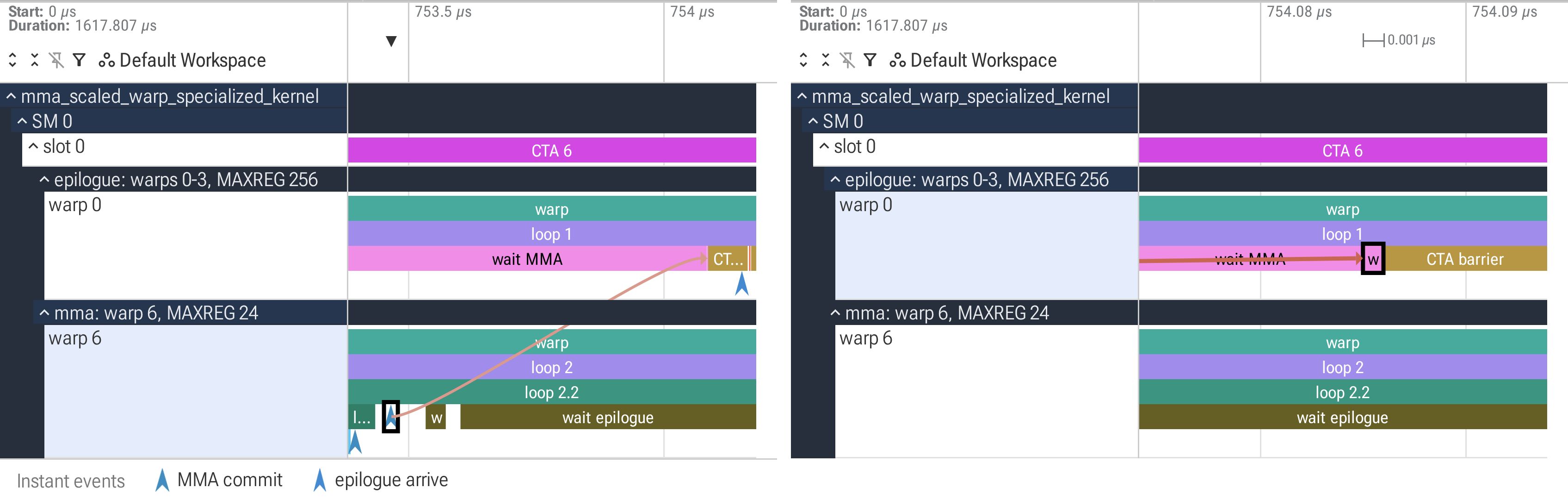}{The epilogue's \texttt{wait MMA} of Figure~\ref{fig:app-reading}. Left: the selected \texttt{MMA commit} and its arrow, which ends where the wait returns. Right: 20\,ns around that point; the wait's last nanosecond (selected) is a slice of its own, and the arrow ends on it.}{fig:app-lastns}

\subsection{Example Timelines}
\label{app:tour}

\paragraph{The whole kernel.}
Figure~\ref{fig:app-panorama} shows the GEMM of \S\ref{sec:gemm-example} at full length.
This view shows whether the hardware stays busy, how the kernel starts and drains, and how many CTAs each SM runs.
All 19 PM series are pinned at the top in their four groups.
Below them, SM~0 runs one CTA after another in its single slot.
Warps 0--7 are the epilogue, warps 8--9 load, and warp 10 issues the MMAs.
Warp 11 does no work: its branch only sets its register limit and waits at two CTA barriers.

\clearpage

\appshot{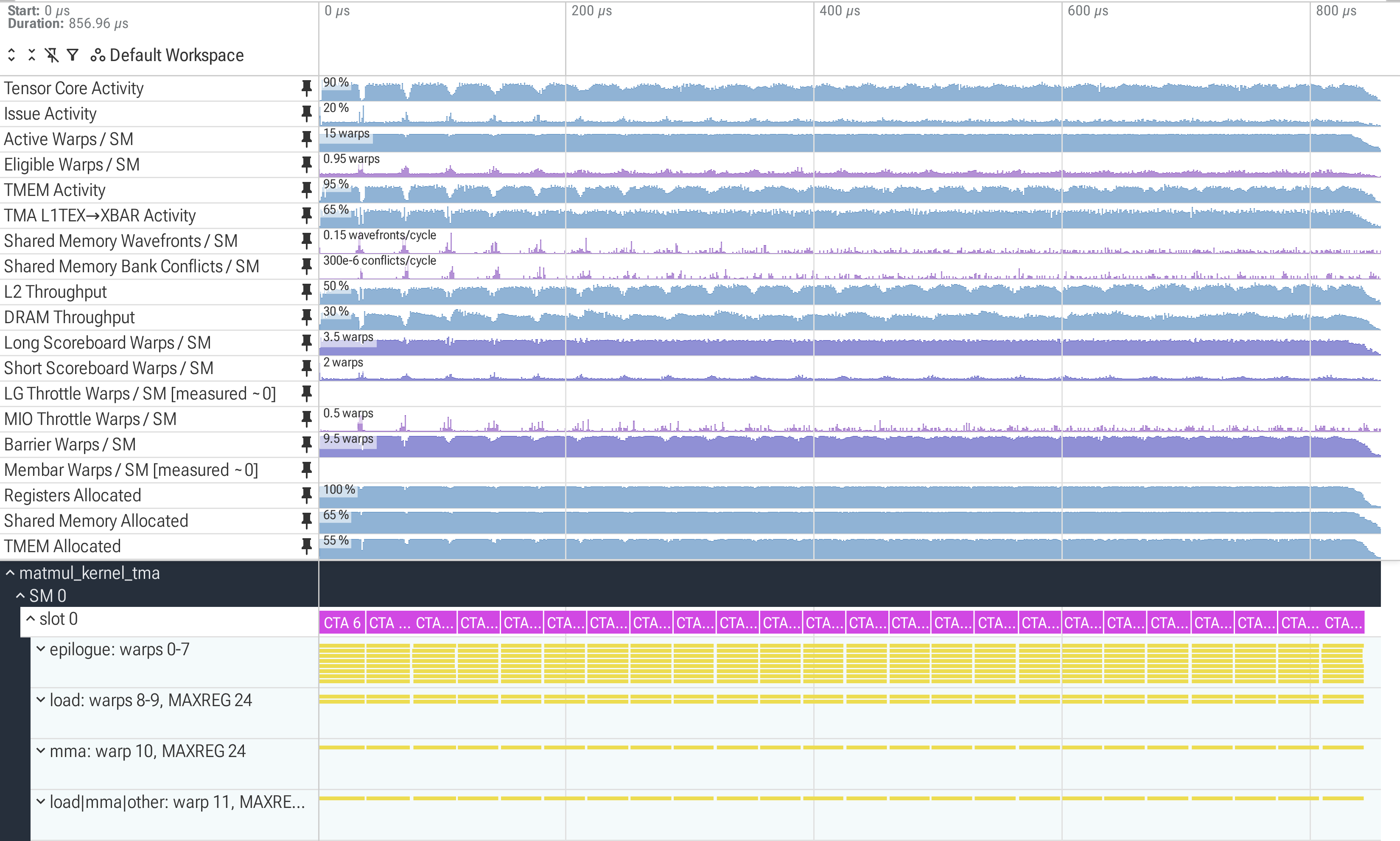}{The whole Triton GEMM of \S\ref{sec:gemm-example}. Top: the 19 PM series. Bottom: SM~0's slot and its four role groups, collapsed.}{fig:app-panorama}

\paragraph{One buffer stage between load and MMA.}
Figure~\ref{fig:app-slot-ring} zooms into the same GEMM and follows one stage of the three-stage shared-memory buffer of Figure~\ref{fig:autotracer} through two uses.
Warp~8 loads the stage's A and B tiles with two \texttt{TMA load}s, while warp~9, the other load warp, only waits with it.
When both copies have landed, \texttt{full[0]} completes and ends the MMA warp's \texttt{wait TMA}, and the arrow starts at the later load.
The MMA warp then issues its MMAs on the stage and commits.
The \texttt{MMA commit} completes \texttt{empty[0]} and ends the \texttt{wait MMA} of both load warps, after which warp~8 loads the stage again.

\appshot{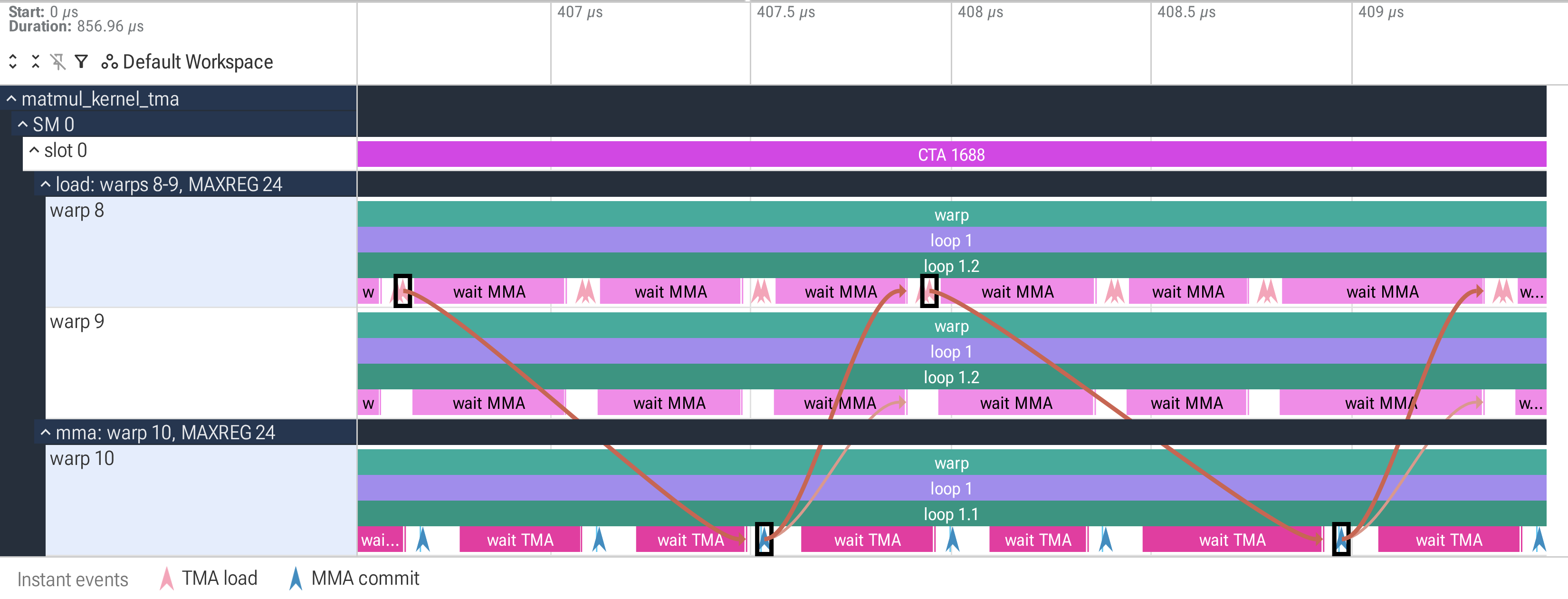}{Two uses of stage~0 of the shared-memory buffer in the Triton GEMM: load warps~8--9 and MMA warp~10 of one CTA on SM~0.}{fig:app-slot-ring}

\clearpage

\paragraph{Reading one dependency.}
Figure~\ref{fig:app-reading} zooms into the production MXFP8 GEMM in Triton, which keeps two accumulators in tensor memory so that the epilogue of one tile overlaps the MMAs of the next.
The selected slice is the last nanosecond of the epilogue's \texttt{wait MMA}, and its arrow comes from the \texttt{MMA commit} that ends the MMA warp's tile.
After the wait the epilogue drains that accumulator with CTA barriers and TMA stores, while the MMA warp works on the next tile in the other accumulator.
Below the timeline are the wait's arguments, explained in Table~\ref{tab:app-args}: \texttt{stage} says which accumulator it waited for, and \texttt{from warp} names the MMA warp ($6 \times 8 + 6$).
Warps 4--5 do no work, like warp~11 in Figure~\ref{fig:app-panorama}.

\makeatletter
\par\medskip\noindent\begin{minipage}{\textwidth}\centering
\includegraphics[width=\textwidth]{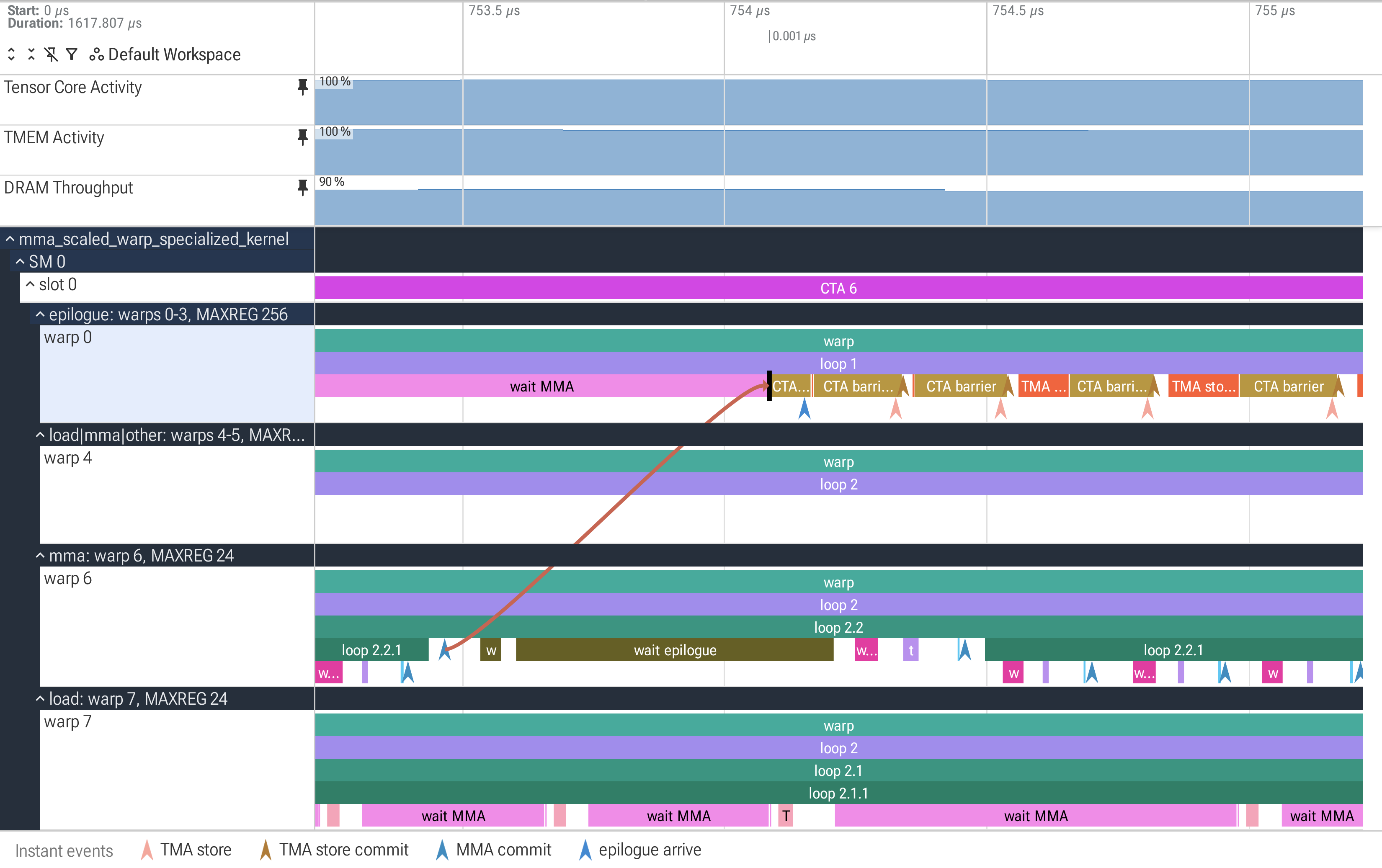}\par
\begin{minipage}[t]{0.30\textwidth}\vspace{0pt}\vspace{36pt}\includegraphics[width=\linewidth]{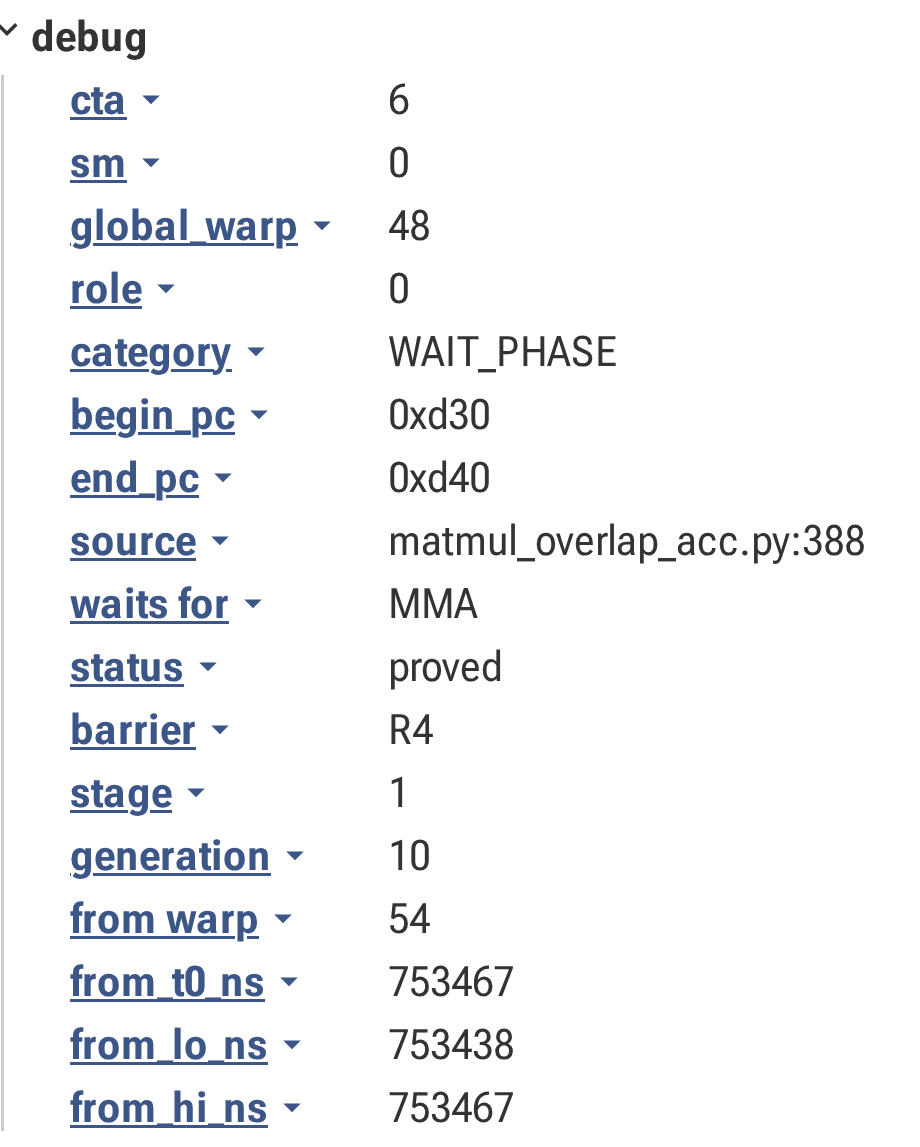}\end{minipage}\hfill
\begin{minipage}[t]{0.66\textwidth}\vspace{0pt}\def\@captype{table}\caption{Arguments of a trace slice. A probe reads the clock after its instruction, not at it, so the trace bounds when an arrow's source issued instead of giving one time.}\label{tab:app-args}
\footnotesize\setlength{\tabcolsep}{4pt}
\begin{tabular}{@{}L{24mm}L{84mm}@{}}
\toprule
Argument & Meaning \\
\midrule
\texttt{cta} & the CTA's index in the grid \\
\texttt{sm} & the SM the CTA ran on \\
\texttt{global\_warp} & CTA index $\times$ warps per CTA $+$ warp index \\
\texttt{role} & the role's index; its track gives the name \\
\texttt{category} & the event (Table~\ref{tab:app-events}) \\
\texttt{begin\_pc} & the SASS address whose probe records the start \\
\texttt{end\_pc} & the SASS address whose probe records the end \\
\texttt{source} & the file and line, from the binary's line table \\
\texttt{waits for} & what completes the wait's barriers (Appendix~\ref{app:barriers}) \\
\texttt{status} & \texttt{proved}, \texttt{observed}, or \texttt{unknown} (Appendix~\ref{app:barriers}) \\
\texttt{barrier} & the barriers it waits on, completes, or arrives on \\
\texttt{stage} & which barrier of a ring this instance used \\
\texttt{generation} & the phase it observed or counts toward (Appendix~\ref{app:deps}) \\
\texttt{from warp} & the global warp of the arrow's source \\
\texttt{from\_t0\_ns} & the start of the arrow's source \\
\texttt{from\_lo\_ns} & the earliest time the source can have issued \\
\texttt{from\_hi\_ns} & the latest time the source can have issued \\
\bottomrule
\end{tabular}\end{minipage}\par
\def\@captype{figure}\caption{The production MXFP8 GEMM in Triton: 2\,\textmu s of CTA~6 on SM~0 at the boundary of two tiles, one warp per role. The last nanosecond of the epilogue's \texttt{wait MMA} is selected; bottom left, its arguments.}\label{fig:app-reading}
\end{minipage}\par\medskip
\makeatother

\clearpage

\paragraph{A closed-source kernel.}
Figure~\ref{fig:app-sdpa} shows cuDNN's flash-attention forward kernel (SDPA), which ships without source.
From its binary \sys{} recovers the roles and their register limits: two softmax warpgroups, a correction warpgroup, and one warp each for MMA, load, epilogue, and scheduling.
The names outline the pipeline.
The correction warp waits on softmax1 and softmax2 in turn, the epilogue warp waits on correction, and the scheduler warp waits on a ring that all six other roles arrive on.
The selected \texttt{MMA commit} holds commits to two barriers, and its arrows end the \texttt{wait MMA} of softmax2 and of correction.

\appshot{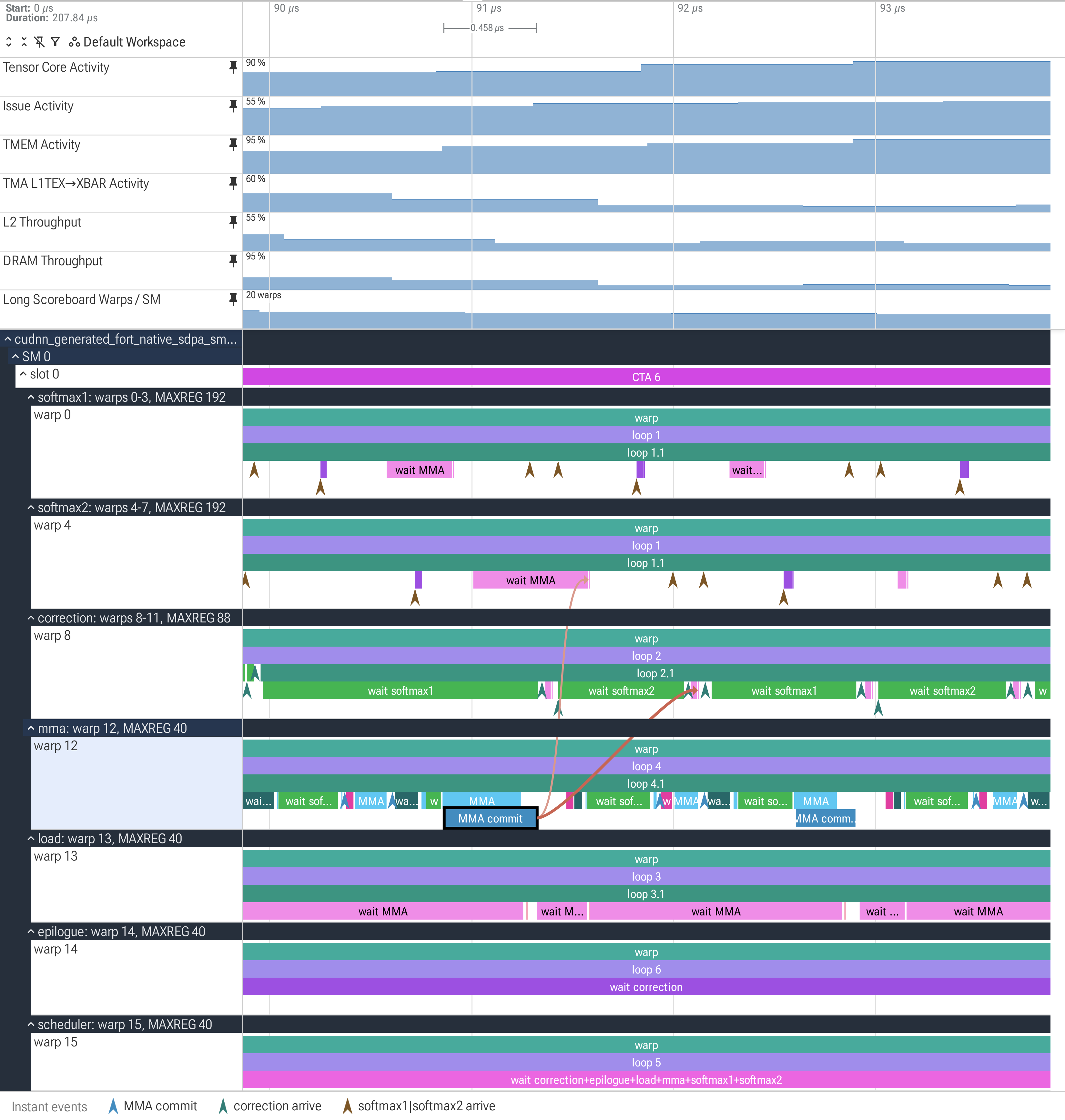}{cuDNN's flash-attention forward (SDPA), closed source: 4\,\textmu s of CTA~6 on SM~0, one warp of each of its seven roles, an \texttt{MMA commit} selected.}{fig:app-sdpa}

\clearpage

\paragraph{CuTe DSL.}
Figure~\ref{fig:app-cute-fmha} shows CUTLASS's FMHA, written in CuTe DSL and compiled by a different compiler than Triton.
Its roles resemble cuDNN's: two softmax warpgroups, correction, MMA, load, and epilogue.
Warp 15 runs no work loop and no classified instruction, so its role is \texttt{idle}.
The selected \texttt{softmax1 arrive} completes phases that softmax2 and the MMA warp wait on, and its arrows end their \texttt{wait softmax1}.
The correction warp's \texttt{wait mma+softmax2} is on a barrier that both of these roles arrive on.

\appshot{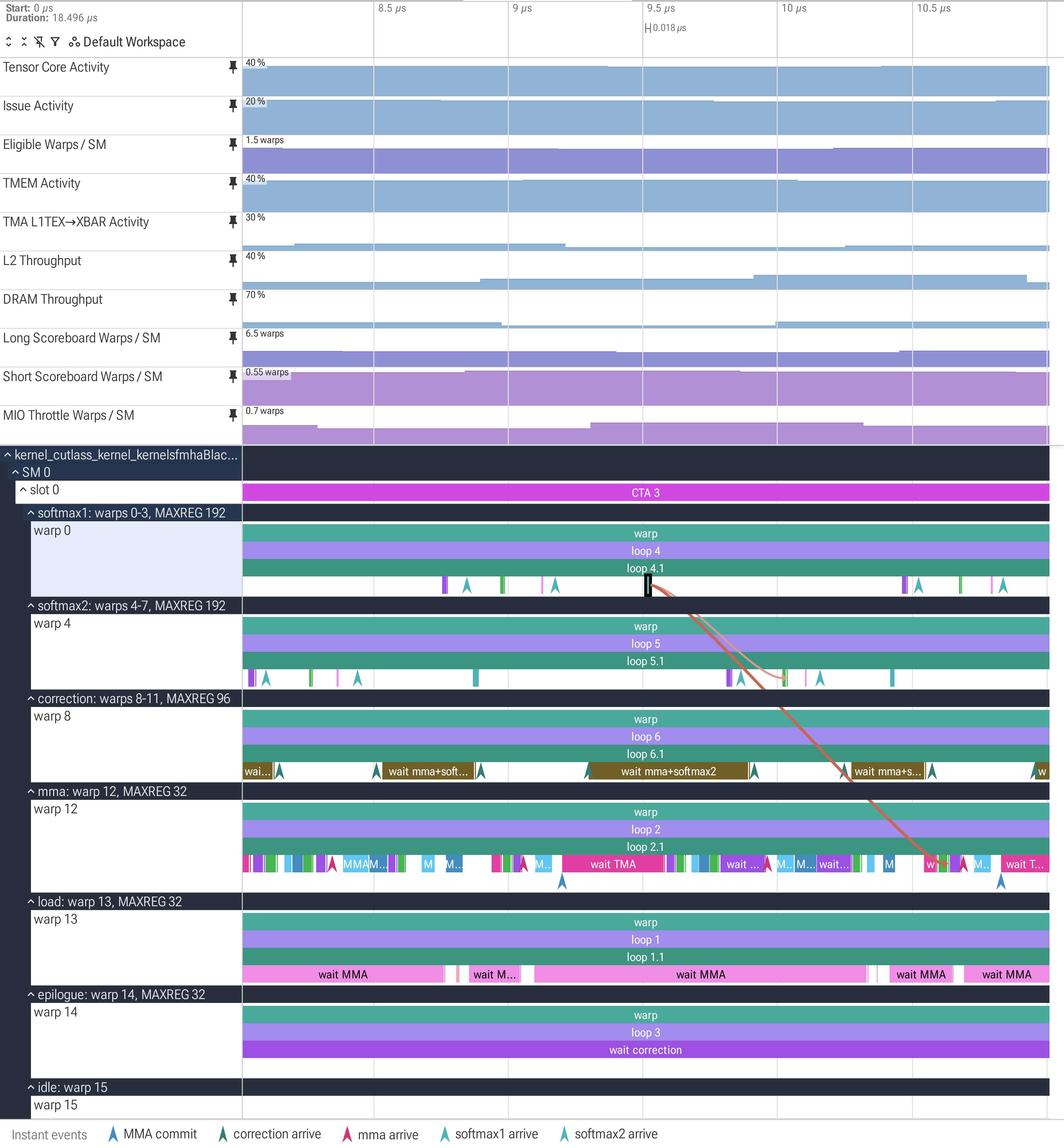}{CUTLASS's FMHA in CuTe DSL: 3\,\textmu s of CTA~3 on SM~0, one warp per role, a \texttt{softmax1 arrive} selected.}{fig:app-cute-fmha}

\clearpage

\paragraph{CUDA C++.}
Figure~\ref{fig:app-nvfp4} shows an NVFP4 GEMM in CUDA C++ that the agent wrote during optimization.
Besides its load, MMA, and epilogue warps it has a \texttt{transform} role of four warps that read shared memory and write shared or tensor memory.
The kernel sets no per-role register limit, so no label shows \texttt{MAXREG}.
The epilogue warps alternate between long waits for the MMAs and bursts of TMA stores.
CTA~7 on SM~1 goes through the same phases at nearly the same times.

\appshot{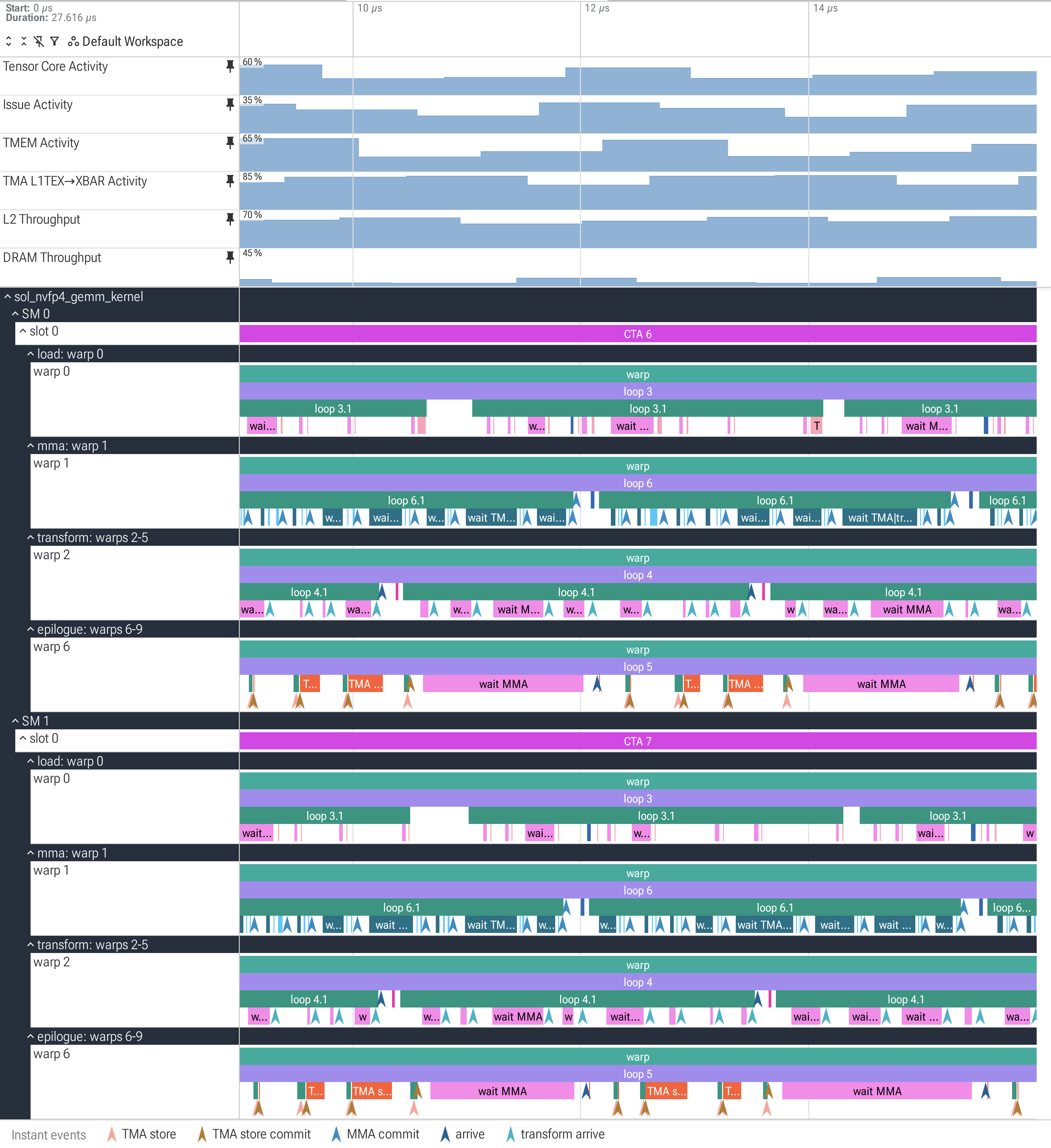}{An NVFP4 GEMM in CUDA C++ written by the agent: 7\,\textmu s of CTA~6 on SM~0 and CTA~7 on SM~1, one warp per role.}{fig:app-nvfp4}

\clearpage

\paragraph{Two CTAs of a cluster.}
Figure~\ref{fig:app-2cta} shows a Triton GEMM whose MMAs span the two CTAs of a cluster.
One CTA of the pair issues each MMA for both, so the MMA warp of CTA~7 runs no MMA loop.
Both epilogues wait for the MMAs and then drain at the same time.
Here the PM series follow the tile boundary: Tensor Core Activity drops in the sample that holds the MMA warp's pause between its two tile loops, and TMA activity peaks in the sample before it.

\appshot{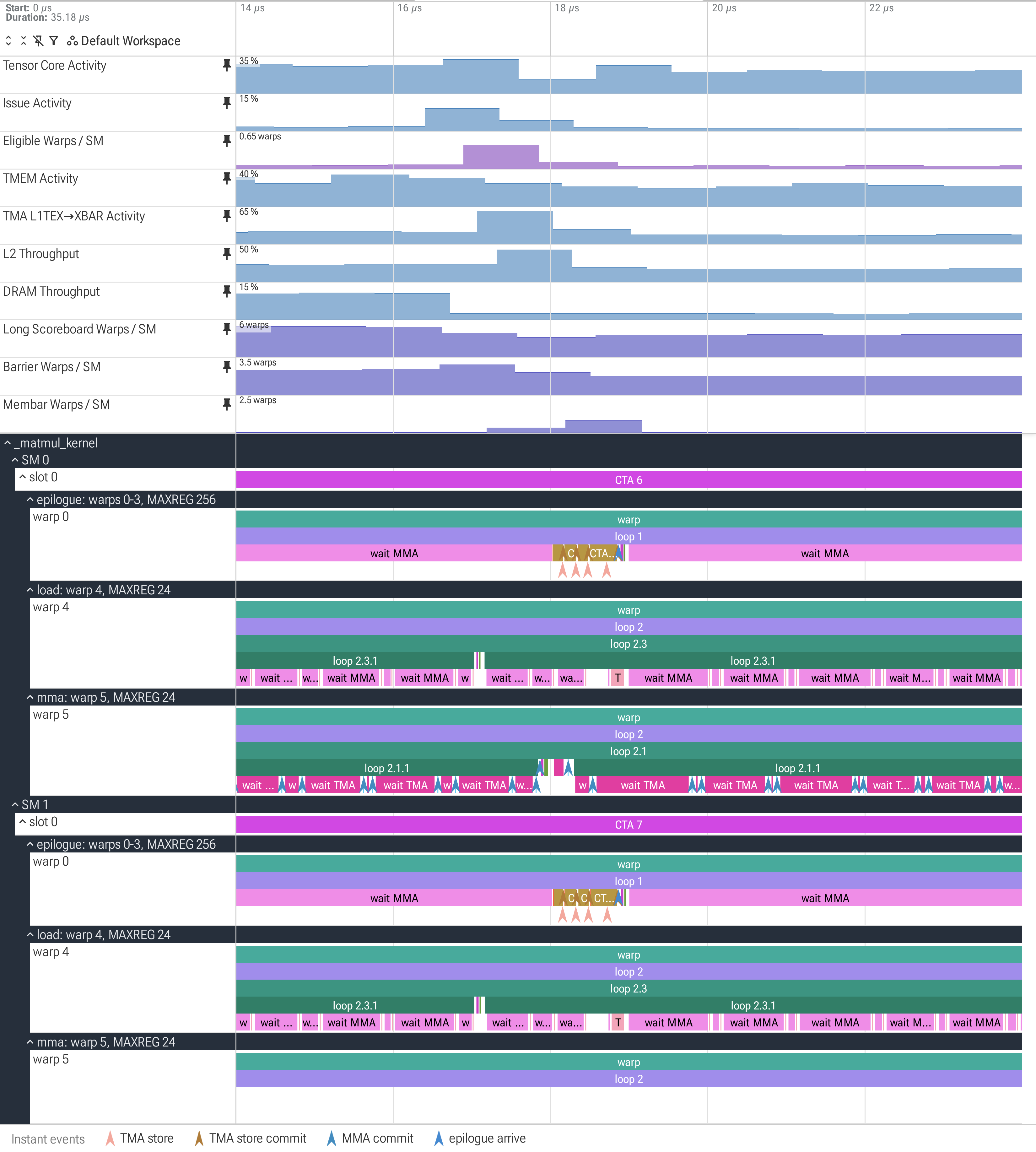}{A Triton GEMM with 2-CTA MMAs: CTA~6 on SM~0 and CTA~7 on SM~1 of one cluster across a tile boundary, the epilogue, load, and MMA roles.}{fig:app-2cta}

\clearpage
\section{How the Evidence Is Derived}
\label{app:derivation}

\subsection{Events}
\label{app:events}

Table~\ref{tab:app-events} lists the events the Probe Plan finds in SASS.
The Lifetime level records only the warp span, and the Pipeline level records every row.
Consecutive instructions of one event in a basic block become one slice, an instant for a single instruction and a span otherwise.
They are split wherever an instruction between them may wait.
A warp span runs from the function's first instruction to its last \texttt{EXIT}, a loop span from the loop's entry to its exit, and a wait span from the first attempt of \texttt{SYNCS.PHASECHK} to the point where success is proved.

\begin{table}[!htbp]
\caption{SASS instructions and the events they become.}
\label{tab:app-events}
\centering\footnotesize
\setlength{\tabcolsep}{4pt}
\begin{tabular}{@{}L{31mm}L{33mm}L{73mm}L{22mm}@{}}
\toprule
Category & Slice name & SASS & Kind \\
\midrule
\multicolumn{4}{@{}l}{\emph{Lifetime level}} \\
\texttt{WARP} & \texttt{warp} & the function's first instruction; every \texttt{EXIT} & span \\
\midrule
\multicolumn{4}{@{}l}{\emph{Pipeline level}} \\
\texttt{LOOP} & \texttt{loop} $n$ & a natural loop with a back edge that is not a retry edge & span \\
\texttt{WAIT\_PHASE} & \texttt{wait} $X$ & \texttt{SYNCS.PHASECHK} (try-wait, test-wait) and the back edges taken when it fails & span \\
\texttt{WAIT\_CHECK} & \texttt{check} $X$ & an in-place \texttt{SYNCS.PHASECHK} on another barrier than its retry region's & span \\
\texttt{TMEM\_ALLOC} & \texttt{tmem alloc} & \texttt{UTCATOMSWS.FIND\_AND\_SET} & span \\
\texttt{TMA\_ISSUE} & \texttt{TMA load} & \texttt{UTMALDG}; \texttt{UBLKCP} and \texttt{UBLKRED} into shared memory & instant or span \\
\texttt{MMA\_ISSUE} & \texttt{MMA} & \texttt{UTCHMMA}, \texttt{UTCIMMA}, \texttt{UTCQMMA}, \texttt{UTCOMMA}, \texttt{UTCMXQMMA} & instant or span \\
\texttt{TMEM\_COPY\_ISSUE} & \texttt{tmem copy} & \texttt{UTCCP}, \texttt{UTCSHIFT} & instant or span \\
\texttt{COMMIT} & \texttt{MMA commit} & \texttt{UTCBAR} & instant or span \\
\texttt{STORE\_ISSUE} & \texttt{TMA store} & \texttt{UTMASTG}, \texttt{UTMAREDG}; \texttt{UBLKCP} and \texttt{UBLKRED} into global memory & instant or span \\
\texttt{STORE\_COMMIT} & \texttt{TMA store commit} & \texttt{UTMACMDFLUSH} that a store issue reaches & instant or span \\
\texttt{STORE\_COMPLETE} & \texttt{TMA store wait} & \texttt{DEPBAR.LE SB0,\,$N$} that a store commit reaches & span \\
\texttt{COPY\_ISSUE} & \texttt{cp.async} & \texttt{LDGSTS} & instant or span \\
\texttt{COPY\_COMMIT} & \texttt{cp.async commit} & \texttt{LDGDEPBAR} & instant or span \\
\texttt{COPY\_COMPLETE} & \texttt{cp.async wait} & \texttt{DEPBAR.LE SB0,\,$N$} that a cp.async commit reaches & span \\
\texttt{MMA\_SYNC} & \texttt{mma.sync} & \texttt{HMMA}, \texttt{IMMA}, \texttt{DMMA} (issue only) & instant or span \\
\texttt{ARRIVE} & $R$ \texttt{arrive} & \texttt{SYNCS.ARRIVE.TRANS64[.RED].A1T0} & instant or span \\
\texttt{CP\_ASYNC\_ARRIVE} & \texttt{cp.async arrive} & \texttt{ARRIVES}; \texttt{SYNCS.ARRIVE.TRANS64.RED.A0T1}; a site the cubin lists as a cp.async arrival & instant or span \\
\texttt{CLC\_QUERY} & \texttt{CLC query} & \texttt{UGETNEXTWORKID} & instant or span \\
\texttt{CLUSTER\_ARRIVE} & \texttt{cluster arrive} & \texttt{UCGABAR\_ARV} & instant or span \\
\texttt{CLUSTER\_WAIT} & \texttt{cluster wait} & \texttt{UCGABAR\_WAIT} & span \\
\texttt{CTA\_BARRIER} & \texttt{CTA barrier} & \texttt{BAR.SYNC}, \texttt{BAR.RED} & span \\
\bottomrule
\end{tabular}
\par\smallskip
\raggedright $X$: what completes the barrier (Appendix~\ref{app:barriers}). $R$: the roles whose warps run the arrival.
\end{table}

\subsection{Warp Roles}
\label{app:roles}

\begin{table}[!htbp]
\caption{Role naming. Top: instruction classes. Bottom: the ordered rules over the classes $C$ of a role's own code and $C_L$ of its part inside work loops, with the work such a role usually does. The first rule that holds names the role.}
\label{tab:app-roles}
\centering\footnotesize
\setlength{\tabcolsep}{4pt}
\begin{tabular}{@{}L{17mm}L{86mm}L{18mm}L{38mm}@{}}
\toprule
Class & SASS & Class & SASS \\
\midrule
\texttt{mma} & \texttt{UTCHMMA}, \texttt{UTCIMMA}, \texttt{UTCQMMA}, \texttt{UTCOMMA}, \texttt{UTCMXQMMA} & \texttt{max} & \texttt{FMNMX}, \texttt{FMNMX3} \\
\texttt{load} & \texttt{UTMALDG}; \texttt{UBLKCP}, \texttt{UBLKRED} into shared memory; \texttt{LDGSTS} & \texttt{ldtm}, \texttt{sttm} & \texttt{LDTM}, \texttt{STTM} \\
\texttt{reduce} & \texttt{UTMAREDG} & \texttt{ldg} & \texttt{LDG} \\
\texttt{store} & \texttt{UTMASTG}; \texttt{UBLKCP}, \texttt{UBLKRED} into global memory; \texttt{STG} & \texttt{lds}, \texttt{sts} & \texttt{LDS}, \texttt{LDSM}; \texttt{STS}, \texttt{STSM} \\
\texttt{mma\_sync} & \texttt{HMMA}, \texttt{IMMA}, \texttt{DMMA} & \texttt{clc} & \texttt{UGETNEXTWORKID} \\
\texttt{ex2} & \texttt{MUFU.EX2} & & \\
\bottomrule
\end{tabular}

\bigskip
\begin{tabular}{@{}rL{19mm}L{53mm}L{85mm}@{}}
\toprule
& Name & Rule & Usual work \\
\midrule
1 & \texttt{mma} & $\texttt{mma} \in C$ & issues the tensor-core MMAs and commits them to barriers \\
2 & \texttt{compute} & $\texttt{mma\_sync} \in C$ & multiplies with \texttt{mma.sync} and accumulates in registers \\
3 & \texttt{softmax} & $\texttt{ldtm}, \texttt{ex2}, \texttt{max} \in C$ & computes the softmax of attention scores held in tensor memory \\
4 & \texttt{load} & $\texttt{load} \in C \lor (\texttt{ldg} \in C \land \texttt{store} \notin C)$ & copies operands to shared memory, or loads indices or page tables \\
5 & \texttt{reduce} & $\texttt{reduce} \in C \land \texttt{store} \notin C$ & reduces into global memory by TMA (dQ in attention backward) \\
6 & \texttt{correction} & $\texttt{ldtm}, \texttt{sttm} \in C_L$ & rescales the output in tensor memory when a row maximum grows \\
7 & \texttt{epilogue} & $\texttt{store} \in C \lor \texttt{ldtm} \in C$ & moves the result from tensor memory to global memory \\
8 & \texttt{scheduler} & $\texttt{clc} \in C$ & asks cluster launch control for the next tile \\
9 & \texttt{transform} & $\texttt{lds} \in C \land (\texttt{sts} \in C \lor \texttt{sttm} \in C)$ & rewrites data from shared memory, e.g.\ to dequantize it \\
10 & \texttt{other} & none above, but a work loop or a class & work no rule identifies, e.g.\ a static scheduler or a barrier relay \\
11 & \texttt{idle} & no work loop and no class & no work \\
\bottomrule
\end{tabular}
\end{table}

\emph{Profiles.}
The static analysis groups warp positions (the CTA's rank in its cluster and the warp's index in the CTA) by the set of work loops they run.
A loop that only repeats a wait or a retry is not a work loop.
A profile is split further at a branch that sends its warps to different work loops, and where its warps wait on different barriers.

\emph{Assignment.}
In the timed run, each warp takes the profile whose work loops and events match what it recorded.
If several match, the role is undecided and named by their names joined with \texttt{|}; a warp that recorded only its warp span is then idle.

\emph{Naming.}
A role is named from the code only its warps run, by the first rule of Table~\ref{tab:app-roles} that holds.
The names are the ones that warp-specialized kernels such as CUTLASS~\cite{cutlass}, FlashAttention~\cite{flashattention3,flashattention4}, and ThunderKittens~\cite{thunderkittens} give their warps.
Roles of one name in a CTA are numbered from 1 (\texttt{softmax1}, \texttt{softmax2}), and a kernel with one role names it \texttt{all}.
The track label adds the role's warps and its \texttt{USETMAXREG} limit where the static analysis determines it.

\subsection{Barriers}
\label{app:barriers}

The static analysis finds each mbarrier object, a shared-memory offset in a CTA, at its initialization.
A \emph{barrier} is the set of objects that one wait instruction may wait on, merged across all waits; objects the dynamic run observed at an unresolved wait are merged the same way.
A barrier with objects at several offsets is a ring and gets the id \texttt{R}$n$, one at a single offset gets \texttt{B}$n$, and the cluster barrier is \texttt{cluster}.
What completes a barrier names it: \texttt{TMA} for a copy, \texttt{MMA} for a tensor-core commit, \texttt{CLC} for a cluster launch control query, \texttt{cp.async} for a cp.async arrival, and the arriving roles for a software arrival.
The \texttt{status} argument is \texttt{proved} if the static analysis names every barrier, \texttt{observed} if the dynamic run supplied some, and \texttt{unknown} if none is named.

\begin{figure}[!htbp]
\centering\footnotesize
\begin{verbatim}
0a60  SYNCS.EXCH.64 URZ, [UR7+0x240a0], UR4  init s=0
0a70  SYNCS.EXCH.64 URZ, [UR7+0x240a8], UR4  init s=1
0a80  SYNCS.EXCH.64 URZ, [UR7+0x240b0], UR4  init s=2
      ...                                   MMA loop
19f0  UIADD3 UR5, UR5, 0x1                  s+1
1a10  UISETP.NE.U32.AND UP0, UR5, 0x3
1a40  USEL UR5, UR5, URZ, UP0               wrap at 3
1a60  ULEA UR15, UR5, UR7, 0x3              UR7+8s
1a90  SYNCS.PHASECHK.TRANS64.TRYWAIT P0,
          [UR15+0x240a0], R3                wait full[s]
      ...
1c20  UIADD3 UR6, UR15, 0x24080             empty[s]
1ca0  UTCBAR [UR6], URZ                     commit
\end{verbatim}
\caption{SASS of ring \texttt{R4} in the GEMM of \S\ref{sec:gemm-example}, abridged: guards and unused operands are dropped and comments added. The register \texttt{R3} holds the parity the wait tests and is unrelated to ring \texttt{R3}.}
\label{fig:app-r4}
\end{figure}

Figure~\ref{fig:app-r4} shows ring \texttt{R4} of the GEMM of \S\ref{sec:gemm-example}.
Three initializations set up objects 8 bytes apart.
In the MMA loop \texttt{UR5} counts the stage $s$ and wraps at 3, and the wait at \texttt{0x1a90} reads \texttt{UR7}$+8s$, so it may wait on any of the three objects and they form one ring.
The load warp's copies complete it, so the wait is named \texttt{wait TMA}.
The commit at \texttt{0x1ca0} arrives on the matching object of a second ring, \texttt{R3}, which the load warp waits on as \texttt{wait MMA}.
The object each execution of the wait names becomes the slice's \texttt{stage}.

\subsection{Dependencies}
\label{app:deps}

A \emph{slot} is one mbarrier of one CTA.
It is reused many times, and each use is a \emph{phase}.
Copies, commits and arrivals complete a phase, and waits wait for it.
A wait only checks whether the phase number is even or odd.
So to draw an arrow the trace must find out which phase a wait saw and what completed that phase.
We draw an arrow only where both answers are certain.

\emph{Which phase a wait saw.}
Phases are numbered $0, 1, 2, \ldots$, and a slot completes one after $N$ arrivals.
A warp's waits on one slot see later and later phases, so its $k$-th wait (from $k=0$) saw at least phase $g = g_0 + k$, its \emph{generation}.
Here $g_0 = -1$ if the slot starts free and $g_0 = 0$ otherwise, and the parity each wait checks must match $g$.
Since a wait checks only parity, it may instead have seen phase $g+2$ or later.
Phase $g+2$ is complete only after $(g+3)N$ arrivals.
So if fewer arrivals can have issued by the time the wait ended, the wait saw phase $g$.

\emph{What completed a phase.}
All records of one SM share one clock, so the arrivals on a slot can be put in order.
The first $N$ belong to phase~0, the next $N$ to phase~1, and so on.
A copy belongs to the phase of the last expect-tx arrival its warp made before it.
We trust the line between phases $k$ and $k+1$ only if a wait that saw phase $k$ ended before anything in phase $k+1$ can have issued.
A phase keeps its contributions only if both of its lines are trusted.

\emph{Admission.}
The counting above is only right if no arrival is missed or put on the wrong slot.
So a slot is used only if its arrivals per phase are known, its arrivals add up to whole phases, every instruction that uses it computes its address from values fixed by the code path (such as a loop counter) and not from loaded data, and every warp that uses it records each use.

\subsection{PM Series and Their Alignment}
\label{app:pm}

Table~\ref{tab:app-pm} lists the 19 PM series of \S\ref{sec:evidence}.
Nsight Compute~\cite{nsight} samples them on the uninstrumented binary.
We normalize every series to the SM level for easier reading.

\begin{table}[!htbp]
\caption{The 19 PM series by group: the track name, the unit after normalization, what Nsight Compute reports (a mean over SMSPs, TPCs, or SMs, or a GPU total), and the base metric, requested as \texttt{pmsampling:} with the suffix of its unit.}
\label{tab:app-pm}
\centering\footnotesize
\setlength{\tabcolsep}{4pt}
\begin{tabular}{@{}L{46mm}L{25mm}L{18mm}>{\raggedright\arraybackslash\scriptsize}p{72mm}@{}}
\toprule
Series & Unit & Reported as & \multicolumn{1}{l@{}}{Base metric} \\
\midrule
\multicolumn{4}{@{}l}{\emph{Compute \& Scheduler}} \\
Tensor Core Activity & \% of peak & -- & \texttt{sm\_\_pipe\_tensor\_cycles\_active\_realtime} \\
Issue Activity & \% of peak & -- & \texttt{smsp\_\_issue\_active} \\
Active Warps / SM & warps & SMSP mean & \texttt{smsp\_\_warps\_active} \\
Eligible Warps / SM & warps & SMSP mean & \texttt{smsp\_\_warps\_eligible} \\
\midrule
\multicolumn{4}{@{}l}{\emph{Memory System}} \\
TMEM Activity & \% of peak & -- & \texttt{sm\_\_mem\_tensor\_cycles\_active\_realtime} \\
TMA L1TEX$\rightarrow$XBAR Activity & \% of peak & -- & \texttt{l1tex\_\_m\_l1tex2xbar\_req\_cycles\_active\_op\_tma} \\
Shared Memory Wavefronts / SM & wavefronts/cycle & GPU total & \texttt{l1tex\_\_data\_pipe\_lsu\_wavefronts\_mem\_shared} \\
Shared Memory Bank Conflicts / SM & conflicts/cycle & GPU total & \texttt{l1tex\_\_data\_bank\_conflicts\_pipe\_lsu\_mem\_shared} \\
L2 Throughput & \% of peak & -- & \texttt{lts\_\_throughput\_realtime} \\
DRAM Throughput & \% of peak & -- & \texttt{dram\_\_throughput} \\
\midrule
\multicolumn{4}{@{}l}{\emph{Warp Stalls}} \\
Long Scoreboard Warps / SM & warps & SMSP mean & \texttt{smsp\_\_warps\_issue\_stalled\_long\_scoreboard} \\
Short Scoreboard Warps / SM & warps & SMSP mean & \texttt{smsp\_\_warps\_issue\_stalled\_short\_scoreboard} \\
LG Throttle Warps / SM & warps & SMSP mean & \texttt{smsp\_\_warps\_issue\_stalled\_lg\_throttle} \\
MIO Throttle Warps / SM & warps & SMSP mean & \texttt{smsp\_\_warps\_issue\_stalled\_mio\_throttle} \\
Barrier Warps / SM & warps & SMSP mean & \texttt{smsp\_\_warps\_issue\_stalled\_barrier} \\
Membar Warps / SM & warps & SMSP mean & \texttt{smsp\_\_warps\_issue\_stalled\_membar} \\
\midrule
\multicolumn{4}{@{}l}{\emph{Runtime Resources}} \\
Registers Allocated & \% of SM capacity & TPC mean & \texttt{tpc\_\_sm\_rf\_registers\_allocated\allowbreak\_shader\_cs\_realtime} \\
Shared Memory Allocated & \% of SM capacity & TPC mean & \texttt{tpc\_\_l1tex\_sram\_bytes\_mem\_untagged\allowbreak\_data\_shared\_allocated\_compute\_realtime} \\
TMEM Allocated & \% of SM capacity & SM mean & \texttt{sm\_\_mem\_tensor\_bytes\_allocated\_realtime} \\
\bottomrule
\end{tabular}
\end{table}

\paragraph{Time alignment.}
The series and the trace come from different runs, so their times must be aligned.
Both runs show how many warps are running per SM over time.
The profiler samples this as Active Warps / SM.
The trace gives it by counting the warps running at each moment and dividing by the number of SMs.
We place every series on the trace with one map $t_{\mathrm{trace}} = a + b\,t_{\mathrm{profile}}$ and fit $a$ and $b$ by least squares so that the two curves agree.
Here $t_{\mathrm{profile}}$ is a sample's time since the kernel started in its replay pass and $t_{\mathrm{trace}}$ is time since the first warp started.

\section{Optimization Prompts}
\label{app:prompts}

Every round of every arm receives one prompt rendered from one template.
The arms differ in one sentence, and the rounds differ only in the round number and the round's paths.
Below is round~1 of the gated-delta-rule problem, with paths abbreviated as $\langle\cdot\rangle$.

\begin{quote}\small\raggedright
Optimize \texttt{sol\_gated\_delta\_rule\_suite} for every workload of the problem on an NVIDIA B200.
Achieve the lowest GPU execution time you can under the provided evaluation conditions while preserving correctness.

This is round 1 of 6 of the optimization loop, and it has 60 minutes from when you start.
When you finish, or when its time is up, the candidate you left in the workspace is evaluated against the current best implementation.
Spend this round on one promising candidate.
Leave the strongest correct version you found as your candidate, and write what you learned and what remains open into the notes for the next round.

\emph{[arm sentence]}

Inputs
\begin{itemize}\itemsep0pt
\item Kernel (yours to change): $\langle$round$\rangle$\nolinkurl{/ws/cand/kernels/sol_gated_delta_rule.py}
\item Problem (fixed: workloads, inputs, reference, tolerance): the file $\langle$problem file$\rangle$, its arguments \texttt{--all-workloads} (every workload it lists is judged)
\item Evaluate: $\langle$round$\rangle$\nolinkurl{/ws/evaluate} {-}{-} checks and times the kernel the way this round's evaluation does; every call is recorded in the history
\item History: $\langle$round$\rangle$\nolinkurl{/inputs/history.md}; every earlier round under $\langle$session$\rangle$\nolinkurl{/rounds}
\item Notes for the next round: $\langle$round$\rangle$\nolinkurl{/ws/notes.md}
\end{itemize}
\end{quote}

The arm sentence is empty for Bare.
For \sys{} it is ``Use the \texttt{warpline-optimize} skill.'', and the message starts with the skill's command \texttt{/warpline-optimize}, which loads the skill before the model reads the prompt.
For KDA the sentence is:

\begin{quote}\small\raggedright
Use the KernelWiki and ncu-report-skill skills and the humanize plugin: this round is one humanize RLCR loop {-}{-} write a draft to \texttt{docs/draft.md}, turn it into \texttt{docs/plan.md} with \texttt{/humanize:gen-plan --input docs/draft.md --output docs/plan.md --direct}, and implement the plan with \texttt{/humanize:start-rlcr-loop docs/plan.md --yolo}, in which Codex reviews every iteration; the round ends when the loop ends or its time is up.
\end{quote}

The framework renders the history file the same way for all arms.
It holds a table of every round (parent, edited level, verdict, speedup over the parent and over the baseline, diff size, number of evaluations, minutes), the chain of accepted rounds that leads to the current best, the gate's measurements for the most recent rounds with every evaluation the agent ran in them, and the previous round's notes.